\documentclass[prx,floatfix,superscriptaddress,nobibnotes,nofootinbib,notitlepage,issn=false,doi=false,longbibliography]{revtex4-2}
\usepackage[left=25mm,right=25mm,top=35mm,columnsep=15pt]{geometry} 
\usepackage{graphicx}
\usepackage{physics}
\usepackage{siunitx}
\usepackage{xcolor}
\usepackage{setspace}
\usepackage{amsthm,amsmath,amssymb}
\usepackage{mathtools}
\usepackage{bm}
\usepackage[T1]{fontenc}
\usepackage{newtxtext, newtxmath, newtxtt}
\usepackage{nicefrac}
\usepackage{booktabs}
\usepackage{multirow}
\usepackage{physics}
 
\usepackage{fancyhdr}

\newcommand{\appropto}{\mathrel{\vcenter{
  \offinterlineskip\halign{\hfil$##$\cr
    \propto\cr\noalign{\kern2pt}\sim\cr\noalign{\kern-2pt}}}}}

\newcommand{\numel}{ \ensuremath{ {N_{ \rm el } }} }
\newcommand{\numnuc}{ \ensuremath{ {N_{ \rm nuc } }} }

\usepackage{dsfont}

\usepackage[pdftex, pdftitle={Article}, pdfauthor={Author}]{hyperref} 

\hypersetup{
    colorlinks,
    linkcolor={red!50!black},
    citecolor={blue!50!black},
    urlcolor={blue!80!black}
}

\usepackage[capitalise]{cleveref}

\newcommand{\uoftcs}{\affiliation{Department of Computer Science, University of Toronto, Toronto, ON M5S 3G4, Canada.}}
\newcommand{\uoftchem}{\affiliation{Department of Chemistry, University of Toronto, Toronto, ON M5S 3H6, Canada.}}
\newcommand{\vvector}{\affiliation{Vector Institute of Artificial Intelligence, Toronto, ON M5G 1M1, Canada.}}
\newcommand{\cifar}{\affiliation{Canadian Institute for Advanced Research, Toronto, ON M5G 1Z8, Canada.}}
\newcommand{\acab}{\affiliation{Acceleration Consortium, Toronto, ON M5S 3H6, Canada.}}
\newcommand{\chemeng}{\affiliation{Department of Chemical Engineering \& Applied Chemistry, University of Toronto, Toronto, ON M5S 3E5, Canada.}}
\newcommand{\matsci}{\affiliation{Department of Materials Science \& Engineering, University of Toronto, Toronto, ON M5S 3E4, Canada.}}
\newcommand{\lebovic}{\affiliation{Lebovic Fellow, Canadian Institute for Advanced Research, Toronto, ON M5S 1M1, Canada.}}
\newcommand{\horizon}{\affiliation{Horizon Quantum Computing, Singapore 138565.}}
\newcommand{\stjude}{\affiliation{Department of Structural Biology, St. Jude Children's Research Hospital, Memphis, TN 38105, USA.}}
\newcommand{\harvardphys}{\affiliation{Department of Physics, Harvard University, Cambridge, MA 02138, USA.}}
\newcommand{\danishcs}{\affiliation{Department of Computer Science, University of Copenhagen, Copenhagen, DK-2100, Denmark.}}
\newcommand{\parischem}{\affiliation{Chimie ParisTech, PSL University, CNRS, Institute for Life and Health Sciences (iCLeHS UMR 8060), 75005 Paris, France.}}

\newcommand{\added}[2][]{#2}
\newcommand{\deleted}[1]{}

\newcommand{\reviseRepl}[2]{#1}
\newcommand{\reviseNew}[1]{#1}

\makeatletter
\def\l@subsubsection#1#2{}
\makeatother

\begin{document}
\pagestyle{plain}
\title{Chemically Motivated Simulation Problems are Efficiently Solvable by a Quantum Computer}
\author{Philipp Schleich}
\email[]{philipps@cs.toronto.edu}
\uoftcs\vvector
\author{Lasse Bj\o rn Kristensen}
\email[]{lakr@di.ku.dk}
\uoftcs \uoftchem \danishcs
\author{Jorge A. Campos-Gonzalez-Angulo}\uoftchem\vvector
\author{Abdulrahman Aldossary}\uoftchem
\author{Davide Avagliano}\uoftchem\parischem
\author{Mohsen Bagherimehrab}\uoftcs
\author{Christoph Gorgulla}\stjude\harvardphys
\author{Joe Fitzsimons}\horizon
\author{Al\'an Aspuru-Guzik}
\email[]{aspuru@utoronto.ca}
\uoftcs\vvector\uoftchem\chemeng\matsci\acab\cifar\lebovic

\begin{abstract}
    Simulating chemical systems is highly sought after and computationally challenging, as the \reviseRepl{number of degrees of freedom}{simulation cost} increases exponentially with the size of the system. Quantum computers have been proposed as a computational means to overcome this bottleneck \added{, thanks to their capability of representing this amount of information efficiently}. \reviseRepl{Most efforts so far have been centered around}{Most recent efforts have been directed towards} determining the ground states of chemical systems. However, hardness results and the lack of theoretical guarantees for efficient heuristics for initial-state generation shed doubt on the feasibility. Here, we propose \reviseRepl{a heuristically guided approach that is based on inherently efficient routines to solve}{an inherently efficient approach for solving} chemical simulation problems, \reviseRepl{ requiring}{ meaning it requires} quantum circuits of size scaling polynomially in relevant system parameters. If a set of assumptions can be satisfied, our approach finds good initial states for dynamics simulation by assembling them in a scattering tree.
    \added{In particular, we investigate a scattering-based state preparation approach within the context of mergo-association.}
    We discuss a variety of quantities of chemical interest that can be measured after the quantum simulation of a process, e.g., a reaction, following its corresponding initial state preparation.
\end{abstract}

\maketitle
\tableofcontents

\section{Introduction}
The idea of using quantum computers for the simulation of quantum systems goes back to the first proposals of quantum computing by Benioff, Feynman, and Manin \cite{benioff_computer_1980,feynman_simulating_1982,manin_computable_1980}. This idea has generated substantial effort toward applying quantum computing to chemical problems, as conventional quantum many-body simulations for chemistry are inherently limited by the curse of dimensionality and constitute a significant portion of current supercomputing usage.
This effort has largely focused on two problems: quantum simulation of dynamics (also known simply as Hamiltonian simulation or quantum simulation in the field of quantum computing) and the ground-state preparation problem.
Quantum simulation \cite{zalka_simulating_1998,wiesner_simulations_1996,abrams_simulation_1997,aharonov_adiabatic_2003,berry_efficient_2007} describes the problem of time-evolving an initial state according to the Schr\"odinger equation under a Hamiltonian.
The relevant Hamiltonians for chemical and most physical processes can be efficiently computed, and the corresponding time evolution, governed by the Schr\"odinger equation, is provably within the computational complexity class BQP on a quantum computer.
This class encompasses decision problems that a quantum computer can solve in polynomial time with a bounded error probability \cite{abrams_simulation_1997,lloyd_universal_1996}.
In contrast, the problem of determining the ground state – formulated as the local Hamiltonian problem in quantum complexity theory – is complete for the class Quantum Merlin-Arthur (QMA) \cite{lloyd_universal_1996,abrams_quantum_1999,kempe_complexity_2005}, a larger class sometimes compared to the classical complexity class NP. 
\added{More recently, it was shown that simulating Schr\"odinger operators  of the form $-\Delta+V$ with some restrictions on $V$ and without particle exchange symmetries, a slight restriction compared to general local Hamiltonians, is also BQP-complete, and the ground-state problem for such operators is StoqMA-hard~\cite{zheng2024computational}; namely Merlin-Arthur given Hamiltonians without sign problem (stoquastic Hamiltonians)~\cite{bravyi2006merlin}.}
As a result, it is not clear whether \reviseRepl{ground state problems}{this problem} can be solved efficiently, even on a quantum computer. 

Ground-state search and Hamiltonian simulation are often used jointly. Simulation algorithms serve as a subroutine to find the ground-state energy (e.g., via phase estimation \cite{abrams_quantum_1999,kitaev_quantum_1995,aspuru-guzik_chemistry_2005,whitfield_simulation_2011}) or in the elucidation of reaction mechanisms described in \cite{reiher_elucidating_2017}), while ground states can be input states for performing quantum dynamics. State-of-the-art implementations of Hamiltonian simulation can be categorized into several classes. One is the class of algorithms based on Trotter-Suzuki formulas \cite{lloyd_universal_1996,childs_theory_2019}, which split the exponential of a sum into products of exponentials. Another class is based on ``qubitization'' \cite{low_hamiltonian_2019,berry_qubitization_2019}, which makes use of quantum signal processing~\cite{motlagh_generalized_2023} to encode the simulation in a quantum walk. Additionally, there are randomized algorithms such as QDRIFT~\cite{campbell_random_2019} and its extensions~\cite{nakaji2023qswift,pocrnic2024composite}. 
Each of these approaches has its advantages in different scenarios~\cite{rajput_hybridized_2022,hagan2023composite}\added{, and methods based on product formulas in particular can make the evolution of perturbative systems, as often present in chemistry and physics, even more efficient~\cite{bagherimehrab2024faster,bosse2024efficient}}. However, as they all lead to polynomially sized circuits for Hamiltonians of constant sparsity, we refrain from going into details here and refer to the relevant literature. A good overview can be found in~\cite{su_fault-tolerant_2021}.

We propose a departure from the current mainstream approach of the quantum computing community to chemistry, moving beyond the 1950s computational chemist’s way of thinking~\cite{lowdin_correlation_1958,lowdin_historical_1995}, which has been shaped by the limitations of classical computing to use the ground-state as the starting point of computations, to a new era. 
With the advent of fault-tolerant quantum computers, dynamical simulation of quantities that a practicing chemist might care about is within reach; after all, most relevant quantum chemistry problems are inherently addressable through dynamical evolution alone, leading to efficient quantum algorithms for these problems.

Our proposed framework employs a limited set of attainable atomic initial states and then dynamically prepares input states for a reaction of interest through a scattering process. 
\reviseNew{Particularly, our main contribution is the approach to prepare molecular states by hierarchically `combining' atomic ones -- here specifically by scattering them. This allows for a state preparation heuristic guided by chemical intuition.}
Then, another dynamical evolution embodies a reaction, and a broad set of relevant quantities can be measured.  
Under certain assumptions on what constitutes relevant initial states, the ensuing algorithm is not limited by the QMA-hardness of preparing ground states and thermal states. Similarly, we can circumvent an orthogonality catastrophe, namely, a vanishing success probability to retrieve a ground state. 
As it has been recently shown~\cite{lee_evaluating_2023}, this effect is why exponential speedups for ground-state energy estimation of molecular Hamiltonians on quantum computers may be hard to attain as the (state-of-the-art) methods considered for state preparation fail to produce reliable overlaps for molecules of increasing size. 
\added{However, a distinction to make is that our methodology is based on heuristical physical intuition, which makes it not directly comparable to methods like QPE from a complexity-theoretic perspective.}

Two observations are at the foundation of our approach. First, the simulation of $k$-local Hamiltonians is achievable by polynomial-size quantum circuits. Hamiltonians that stem from chemical problems are 2-local due to the nature of the Coulomb interaction and, thus, also have finite locality when represented as quantum operations on qubits. Including photons to simulate light-matter interactions in the computation increases the maximum support of operations but not beyond a constant factor. The second observation is that we aim to simulate processes corresponding to experiments that can be performed in a lab in finite time, i.e., problems that are, in some sense, experimentally verifiable. Molecular ground states, as viewed from the perspective of computational chemistry, which implies a frozen molecular geometry at absolute-zero temperature, typically do not represent a system's thermal equilibrium state and hence do not belong to this class. 
We propose to simulate the process of producing reactants with a hierarchical scattering process. First, we prepare the ground states of atoms by quantum phase estimation -- this does not fall under the orthogonality catastrophe as atoms are finite-sized and as long as we stay within such where efficient heuristics for initial state preparation are available -- and then use a simple scattering process to produce molecular reactants.  Our method integrates artificial potentials and photonic fields to induce the success of scatterings, leading to a lower-bounded probability of success. \added{Specifically, mergo-association as discussed in \cref{sec:scattering-via-mergo-etc} is a promising avenue to realize this.} This means that a fixed number of repetitions of a weak measurement scheme -- see \cref{subsec:meas-oracles-main-text} -- to herald success will suffice to ensure the success of intermediary scattering processes. 
Thus, meaningful molecular input states, which do not need to be ground states, are efficiently prepared.
Our framework facilitates the modeling of complex chemical reactions by hierarchically operating the scattering with $N$ atoms to create $M$ reactants, which can then undergo a quantum simulation corresponding to a reaction, followed by measurements of reaction rates and time correlation functions.

\section{Hamiltonian Simulation and Complexity Considerations}
We continue by discussing relevant complexity classes for the problems we consider and the efficiency of their solutions.
The complexity class BQP (bounded-error quantum polynomial-time) is oftentimes considered the quantum generalization of the complexity class P (polynomial time), or, more precisely, its probabilistic extension BPP (bounded-error probabilistic polynomial-time). Polynomial complexity is usually considered efficient as the increase in cost when scaling relevant parameters is somewhat moderate.
Problems from QMA (Quantum Merlin-Arthur), in contrast, can be considered ``hard''; QMA is the quantum analog of NP~(non-deterministic polynomial time) or, more precisely, the probabilistic class MA (Merlin-Arthur). QMA describes promise problems with solutions that can be verified in polynomial time with bounded error probability. However, there is no guarantee of efficiency in finding their solution.~\cite{ambainis_physical_2014,watson_complexity_2021,bausch_uncomputability_2021}
Our argument is that a large class of chemically relevant phenomena can be addressed by algorithms solving efficiently solvable problems. More precisely, the intuitive argument we make is that phenomena that occur in finite time in a chemical laboratory can be simulated in finite time. The gap between time in nature and time on a quantum computer is only polynomially sized, which is why we relate it to BQP. Systems in the world around us during chemical reactions typically do not cool down to the ground state; thus, it would not be part of the chemically relevant quantities we consider. In fact, it was recently shown that local minima are more favorable to reach in quantum systems than ground states~\cite{chen2023local}. 

\begin{figure}[h!]
    \centering
    \includegraphics[width=.7\linewidth]{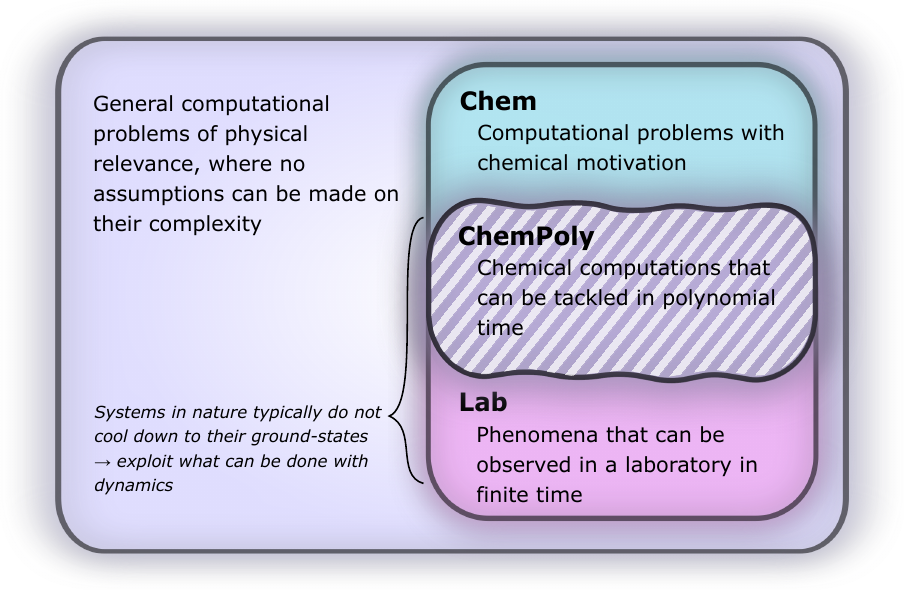}
    
    \caption{Complexity of solving chemical problems. We target the set of computational problems, \textsc{ChemPoly}, consisting of problems within chemistry with polynomial complexity when solved on a quantum computer. These problems have unknown overlap with those that correspond to observables in a laboratory; the sets \textsc{Lab} and \textsc{ChemPoly} may also coincide.}
    \label{fig:complexity_overview}
\end{figure}

Here, we can draw a distinction between polynomially-sized problems, such as those from BQP, and QMA-hard problems, such as finding the ground state. 
In this way, we envision a fundamental change in the way chemistry problems on quantum computers are typically approached, and we display this in \cref{fig:complexity_overview}. We denote by \textsc{Chem} the set of computational problems of chemical relevance – including phenomena like ground states. Note that these are not necessarily decision problems; our argumentation is based on circuit fragments that stem from known complexity results rather than decision problems and the language of complexity theory. We call \textsc{ChemPoly} to the focal set of problems within this work, namely, problems of relevance to chemistry and that have polynomial complexity. 
Due to the BQP-completeness of Hamiltonian simulation~\cite{feynman_quantum_1985}, perhaps one can reduce polynomially-sized problems to ones from the set of problems with chemical relevance. 
Irrespective of this point, our present study highlights a wide set of chemically interesting problems involving simulation as a subroutine yet avoiding subroutines and decision problems known to be QMA-hard. 
Many approaches in chemistry that involve searching for the ground state are \deleted{either} QMA-\added{, StoqMA} or NP-hard~\cite{schuch_computational_2009,whitfield_computational_2012,ogorman_intractability_2022,zheng2024computational}; examples of tasks that fall into this category are finding the universal functional in density functional theory or the Hartree-Fock problem. Significant progress was made on solving the latter numerically, to the point that it is mostly seen as ``solved'' despite the formal hardness. This view results in hope about tackling the ground-state problem as well. Yet still, at this point, we seem far from similar success on chemistry ground-state search, and the biggest hope for solving practical chemical problems lies in dynamics.

To represent chemical systems, we choose molecular Hamiltonians that are not necessarily restricted to the Born-Oppenheimer (BO) approximation. Non-BO was previously explored in the split-operator formalism on quantum computers, where kinetic and potential dynamics are factored~\cite{kassal_polynomial-time_2008,chan_grid-based_2023}. While it was argued in~\cite{kassal_polynomial-time_2008} that the non-BO approach is more efficient than implementing the BO procedure, there have been recent advances that render BO more efficient thanks to a fully coherent implementation~\cite{simon2024improved}. Hamiltonians occurring in chemistry are composed by the operators outlaid in \cref{tab:ham-components}, from which locality is an obvious consequence, as $k$-locality follows from the 2-body nature of the interactions.

\begin{table}[ht!]
    \centering
    \begin{tabular}{l >{\centering\arraybackslash} m{8cm}}   
    \toprule
         Charged particle kinetic energy&  $\displaystyle \frac{\mathbf{p}^2}{2m}$\\  
         Photon energy& 
    $\displaystyle \omega \left( a^\dagger a + \frac{1}{2} \right)$\\  
    Interparticle potential&$\displaystyle \frac{qq'}{\lvert \mathbf{r}-\mathbf{r}' \rvert }$\\  
    Photon-particle interaction&$\displaystyle \frac{q}{m} \mathbf{A}(\mathbf{r}) \left[ \frac{q}{2m} \mathbf{A}(\mathbf{r}) - \mathbf{p} \right]$\\
    \bottomrule
    \end{tabular}
    \caption{Hamiltonian components for molecular systems interacting with photons. For a given particle with mass $m$ and charge $q$, $\mathbf{p}$ and $\mathbf{r}$ are, respectively, the corresponding momentum and position operators. For a photon with frequency $\omega$, $a$ and $a^\dagger$ are the corresponding annihilation and creation operators. $\mathbf{A}(\mathbf{r}) \propto \mathbf{c}(\mathbf{k},\mathbf{r})a+\mathbf{c}(\mathbf{k},\mathbf{r})^* a^\dagger $, where $\mathbf{k}$ is the photon wave-vector, and $\mathbf{c}(\mathbf{k},\mathbf{r})$ is a polarization vector, is the vector potential corresponding to the photon field.}
    \label{tab:ham-components}
\end{table}

Our framework is independent of the choice of basis used to represent the Hamiltonian numerically and the choice between first and second quantization. Asymptotically, a first-quantized representation tends to be more efficient for fault-tolerant quantum algorithms with an abundant number of logical qubits, as the number of required qubits grows linearly with the number of particles and logarithmically with the number of basis functions (since one stores the basis information for every particle), as opposed to the linear dependence in the number of basis functions for second quantization, where one stores occupation for each basis function.\added{~\cite{su_fault-tolerant_2021,babbush_exponentially_2016,kivlichan_quantum_2018,babbush_encoding_2018,babbush_low-depth_2018,mcclean2020discontinuous}}
\added{In this work, we restrict ourselves to  the first-quantized representation for concrete examples.}
For a circuit to have polynomially-sized complexity, it is further necessary that the operator norm of the simulated Hamiltonian, or rather, its individual terms acting on at most $k$ qubits, grow at most polynomially in the number of qubits used to represent them~\cite{aharonov_adiabatic_2003,berry_efficient_2007,lloyd_universal_1996}. 
Energy is an extensive thermodynamic property, so it grows roughly linearly by increasing the system size (i.e., the number of particles). Therefore, the thermodynamic relation between the amount of matter and internal energy is linear.

\section{Computational Framework}
\begin{figure}[h!]
    \centering
    \includegraphics[width=\linewidth]{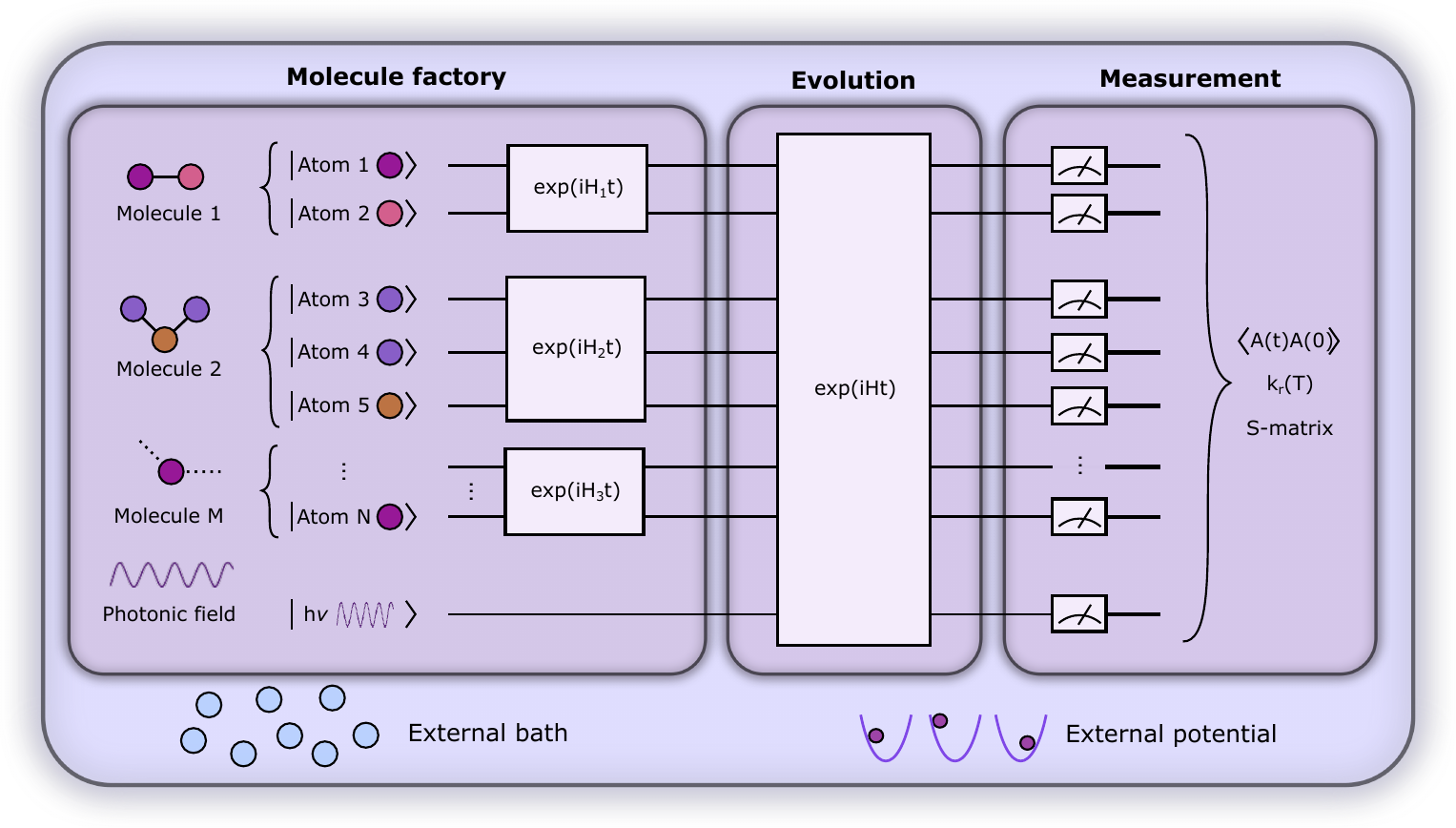}
    \caption{Computational framework. Our simulation framework can be separated into a state preparation procedure (“molecule factory”), the evolution of the reaction of interest, and a measurement step that extracts usable information. The molecule factory prepares a set of molecular input states for the reaction, which may resemble thermal or ground states. These states are produced in a tree-like fashion equipped with a weak measurement scheme to ensure that the target states are prepared with sufficient probability in a heralded way. A photonic field serves as a source of energy for reactions, and an external bath, either explicit or implicit, serves as an energy sink. Furthermore, we utilize external potentials in the spirit of optical tweezers tailored to the different Hamiltonians along the procedure to ensure sufficient success probability and to control positions in space. 
}
    \label{fig:framework}
\end{figure}

In what follows, we describe the computational framework in more detail\added{ -- i.e., a dynamics-based state-preparation scheme that serves as input for a ``main'' quantum dynamics routine, preceding measurement}. 
\deleted{We discuss how to represent systems, their evolution, and chemically interesting observables that can be framed as measurements.}
Our overall idea is based on the experiment in~\cite{liu_building_2018}, where molecules were “built” using two atoms with finite success probability\added{, as well as a mergo-association scheme that merges atoms, e.g. Rb and Cs, using a trap potential~\cite{ruttley2023formation,bird2023making,bird2024makingmoleculesmergoassociationrole}}.
The chemical dynamics we aim to simulate to extract chemical properties requires a sensible initial state for the time evolution. This state must reproduce a natural one faithfully extract the desired properties. \reviseRepl{Depending on the process under consideration, good candidates could be found among ground-states, or perturbed ground-states, such as those obtained by laser excitation.}{A ground state would be an obvious candidate.} 
However, as previously argued, general molecular ground states are hard to access. Nevertheless, preparing the ground states of atoms is feasible – the lighter the atom, the easier – and can be done efficiently as constant overlap for heuristic input states is expected~\cite{lee_evaluating_2023}. For this reason, we propose to follow a hierarchical approach, as outlined in \cref{fig:framework}. All processes generating the reactants and products are carried out through Hamiltonian simulation with local Hamiltonians, including molecular and external field components. 
After the using this technique to generate initial states, we aim to measure observables, e.g., reaction rates according to the scheme in~\cite{kassal_polynomial-time_2008,pedernales_efficient_2014} or auto-correlation functions~\cite{kokcu_linear_2023}.
 
We start with a state representing $N$ atoms, all of which are assumed to be in their ground states or another state of interest $\rho_i^{(\text{atom})}$, such as a thermal state. The preparation of these states can be achieved by existing algorithms for ground-state~\cite{abrams_quantum_1999,aspuru-guzik_simulated_2005,ding2024single,berry2024rapid,huggins2024efficient} and thermal-state preparation\reviseRepl{~\cite{chen2023quantum,hagan2025thermodynamic}}{~\cite{chen2023quantum}} followed by amplitude amplification to enhance the probability of obtaining the desired state.
\added{For generality, we represent both pure and mixed states using density-matrix notation.}
Since the atoms are all finitely sized, and we can prepare an initial state with a significant overlap with the desired state for each atom, any amplitude amplification costs are constant with respect to overall system size. The overall cost to prepare the atomic initial states is $O(N\; \text{poly}(\frac{1}{\varepsilon}))$, with $\varepsilon$ being the accuracy in preparing the atomic states. Using these atoms encoded as a state $\rho_1^{(\text{atom})}\rho_2^{(\text{atom})}\cdots\rho_N^{(\text{atom})}$, we seek to create $M$ molecular reactants $\rho_1^{(\text{mol})}\rho_2^{(\text{mol})}\cdots\rho_M^{(\text{mol})}$.

Based on these initial states, we next prepare the reactants by a scattering process mimicking a physical experiment. In other words, we jointly evolve a set of atoms meant to form a reactant molecule.
We discuss the modeling and treatment of a bath that allows exchanging energy with an environment further below. 

Consider one of the reactant molecules to be prepared. We combine the constituent atoms by successively scattering in a tree-like manner, as in~\cref{fig:framework}, until the desired molecular state is attained. In the worst case, each molecule is generated by combining only two participants at a time, leading to a binary scattering tree, and thus an overall number of scatterings that is quasi-linear in the number of input atoms. 
It is essential to ensure a high overlap with the desired intermediate state at each level in this procedure. 
Otherwise, the overlap would decrease at each combination step. For instance, with an initial overlap of $(1-\delta)$, the overlap would drop exponentially to $(1-\delta)^n$ after $n$ steps. 
A seemingly obvious choice for mitigating this problem would be to use oblivious amplitude amplification~\cite{berry_exponential_2014}. 
However, in our case, the open-system character of the simulation and the fact that the abstract angle to be amplified is not independent of the (unknown) input state are major roadblocks for this approach, which we leave up to future research. 
Instead, we propose the following approach towards bounded success probability.
\reviseRepl{Inspired by}{Following} the nanoreactor approach in~\cite{wang_discovering_2014} \added{and the mergo-association scheme from~\cite{ruttley2023formation,bird2023making,bird2024makingmoleculesmergoassociationrole}}, we introduce artificial potentials that confine the products to be combined in each scattering step, as well as an additional photonic field as energy source and a bath as energy sink. 
\added{Following the procedure we propose in \cref{sec:merging-mols-by-mergoass}, mergo-association shows more promise in terms of suitability for simulation on the molecular scale at this point as we can confine individual, small systems without the need for large 
ensembles in high-pressure and high-temperature regimes.}
Using this framework, we show that, heuristically, we can expect a constant lower bound $P$ on the probability of success for certain types of bonds for the chemical formation of reactants. 
\added{Such a lower bound is an important assumption in the procedure and also highlights its heuristic nature. For the example of mergo-association, we show that the existence of a constant lower bound on the success probability for certain types of bonds, such as covalent bonds, is reasonable; cf. \cref{sec:merging-mols-by-mergoass}.
The intuition we observe there is that given a certain order of magnitude of bonding energy (say, the typical regime of covalent bonds of approximately $30-180~\frac{\rm kcal}{\rm mol}$), then a requirement of simulating potentials of similar magnitude may lead to a substantial constant factor in the simulation cost, but the scaling with respect to system size of this cost would heuristically be only polynomial in the combined nuclear masses, not exponential.
}

\begin{figure}[h!]
    \centering
    \includegraphics[width=\linewidth]{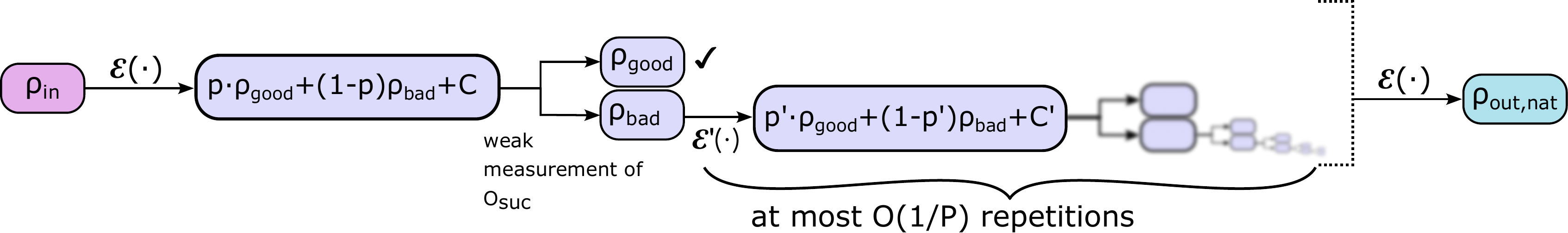}
    \caption{Procedure of a single scattering step. Evolution through the simulation channel $\mathcal{E}(\cdot)$ produces states overlapping successful and not successful spaces. 
    To project the state onto one of these subspaces, a weak measurement is performed, yielding either success or failure. 
    In the case of success, proceed and potentially apply another step of time evolution to ensure the state represents a natural state. If unsuccessful, perform an additional, possibly modified, time evolution, which again produces overlap in the successful subspace, then measure again. Repeat this until success, the expected number of repetitions scaling inversely in the lower bound $P$ on the success probability per step. Success can be quantified by a weak measurement of an observable ${\mathbb O}_\text{suc}$ that, e.g., signifies the success of forming a bond by capturing spatial proximity.}
    \label{fig:scattering-step}
\end{figure}

Suppose the scattering is organized as a \deleted{at least} binary\added{ (or higher-order, ternary, \ldots)} tree. For each node in the tree, there is a simulation channel $\mathcal{E}(\rho)$, parametrized by the subsystem size, dissipation model (bath), and confinement properties (artificial potentials). Then, we may assume that this channel produces a state of the kind 
\begin{equation}
 \mathcal{E}(\rho_0)=p_0 \rho_{\text{suc}}+(1-p_0)\rho_{\neg \text{suc}}+C   ,
 \label{eq:first_state_bipartition}
\end{equation}
with given input $\rho_0$ and probability of success $p_0$; $C$ describes any potential coherence between the subspace of successful and unsuccessful scatterings\added{ and comes from overlap terms such as ${\rm tr}[\rho_{\rm suc}^\dagger\rho_{\neg\rm suc}]$}. 
We may similarly assume we have an observable ${\mathbb O}_\text{suc}$ that allows us to distinguish between the subspaces spanned by $\rho_{\rm suc}$ and $\rho_{\neg \rm suc}$ by weak measurements, which enables the heralding of the scattering success and projecting the state into the successful subspace. The presence of a coherence $C$ does not influence measurement outcomes of $\mathrm{tr}[{\mathbb O}_{\text{suc}} \mathcal{E}(\rho_0) ]$\added{; see \cref{subsec:oracle_on_superpositions}}. Using these two building blocks, we will now outline the simulation strategy of each node, as depicted in \cref{fig:scattering-step}.
The first thing to note is that said measurement of ${\mathbb O}_\text{suc}$, even if carried out weakly, could disturb the state such that it does not describe a realistic state encountered in nature anymore after the measurement. However, we assume that the simulation channel $\mathcal{E}(\cdot)$, such as the one devised below through mergo-association, does resemble a process in nature. Thus, it tends to make states decay towards such physically meaningful states, and another application of the channel will, therefore, map us back into a state resembling those in nature,
\begin{equation}\label{eq:success-channel}
\mathcal{E}(\rho_\text{suc})=p_\text{suc} \rho_\text{suc}^\text{nat}+p_{\neg\text{suc}} \rho_{\neg\text{suc}}^\text{nat}+C^\text{nat}.  
\end{equation}
If the molecule is unstable, but the reacted state is desired, we may simply skip this reapplication and proceed. Hence, if measuring ${\mathbb O}_\text{suc}$ yields a positive outcome, generating meaningful states with a high success rate seems plausible. 
Further, we can use the following strategy to ensure success even if ${\mathbb O}_\text{suc}$ shows that scattering was unsuccessful. Although we are projected into the unsuccessful subspace, the failure is heralded. Therefore, we can apply another simulation channel $\mathcal{E}'(\cdot)$, which may be slightly modified from $\mathcal{E}(\cdot)$ (e.g., stronger confinement or dissipation), leading to 
\begin{equation}\label{eq:non-success-channel}
    \mathcal{E}(\rho_{\neg \text{suc}})=p_1 \rho_{\text{suc}}'+(1-p_1)\rho_{\neg \text{suc}}'+C'.
\end{equation}
This assumption is similar to that in~\cite{ding2024single} in that we expect a redistribution of probability toward the state of interest. We can then redo the measurement and iterate between measurements of ${\mathbb O}_{\text{suc}}$ and (progressively more) modified simulation channels until success occurs.         
To summarize, we can use the above strategy to create a tree-like sequence of scatterings so that, at each step, the success probability at each of the nodes in the tree is bounded by, say, $1\ge P>0$. As we can herald success, at most $O(1/P)$ repetitions are required per node. Since $P$ is promised to be fixed, the number of repetitions per node is $O(1)$. Furthermore, since failure does not require a restart from the leaves of the tree, the complexities of individual nodes straightforwardly add up. Hence, the number of repetitions grows linearly with the number of nodes in the tree, which, with $N$ initial constituents, goes at most as $O(N \log(N))$.
\added{In \cref{subsec:meas-oracles-main-text,app:meas-oracles-appendix}, we provide a more detailed discussion of the construction of the weak measurements.}

We leverage parallels to optical tweezers~\cite{neto_theory_2000,molloy_lights_2002} or molecular beams~\cite{casavecchia_chemical_2000,van_de_meerakker_taming_2008} that are used in physical experiments to engineer the aforementioned artificial potentials to boost the success probability of scatterings.
\added{Before we go into details regarding a specific instance in terms of mergo-association in~\cref{sec:merging-mols-by-mergoass},  we first outline more general approaches here.}
\reviseRepl{
Molecular beams offer well-established success rates, though they require an abundance of particles and, hence, an abundance of available qubits.  
On the other hand, traps via artificial potentials – e.g., adding to the Hamiltonian a harmonic potential term confined to a specific region as done in \cref{eq:trapped-adiab-hamiltonian} – can considerably boost success without high additional space requirement.}{
One could use an abundance of particles, requiring an abundance of available qubits, to use well-established success rates of molecular beams. On the other hand, if we trap particles with an artificial potential – e.g., adding to the Hamiltonian a harmonic potential term confined to a specific region – we can considerably boost success without high additional cost in necessary space.} 
Many of the applications we will discuss benefit from reactant states in the form resembling nuclear wave packets. To that end, consider the input states for the scattering process to be prepared as such. Then, the said process supported by artificial potentials would, in most physical cases, largely preserve the locality in configuration space and maintain the wave-packet like nature of the states.   
\deleted{To induce the interaction of two reactants, we conceptually follow the experimental realization in~\cite{liu_building_2018}.} 
\reviseRepl{
At this point, one may choose to induce a reaction by explicitly modeling a photonic reservoir to excite the reactants ~\cite{cohentannoudji_photons_1997}. Additionally, an explicit register for a bath can be used to absorb excess energy once the reaction has occurred, allowing the products to relax. 
Alternatively, instead of explicitly tracing out a subsystem here, post-selection on a `successful' reaction, as described later in \cref{subsec:meas-oracles-main-text}, can serve to model energy moving out of the system. Finally, energy dissipation can be implemented using Markovian open-system simulation methods, as discussed in more detail below.
}{
Photons are then brought into play to excite the reactants so that the reaction can occur~\cite{cohentannoudji_photons_1997}. A bath can then absorb excess energy once the reaction has occurred, allowing the products to relax. 
}

\reviseRepl{
Beyond the scattering approach, one could think of using a molecular Hartree-Fock (HF) state as an alternative heuristic for an initial state for the reaction dynamics under investigation. 
One can expect the efficacy of this approach to be limited to cases where the Hartree-Fock state as input to a simulation channel 
is close enough to the manifold of physically meaningful states so that convergence to a desired state as input to a reaction happens in controllable time, such as in systems with low static correlation. 
While such systems tend to be amenable to classical treatment, they may be candidates for early experiments of our approach on quantum devices, as it is likely that using an HF initial state has considerably lower constant factors than the scattering approach.
In contrast, the scattering approach will apply to more general states (such as those where Hartree-Fock does not provide a sufficient heuristic). 
}{
One can also think of avoiding the scattering approach by using a molecular Hartree-Fock (HF) state as the initial state, with subsequent mapping of the single-particle electronic wavefunctions to plane waves by a Fourier transform, and the nuclei can be initialized as wave packets in nuclear configuration space. 
Even though it is known that preparing the HF state is not efficient in general~\cite{whitfield_computational_2012}, practical implementations throughout the Roothaan-Hall equations usually scale polynomially with system size (e.g., if the number of iterations is bounded by a constant, which can be chosen to be arbitrarily large)~\cite{gulania_limitations_2021}. 
Then, we can expect that a Hartree-Fock state as input to a simulation channel $\mathcal{E}(\cdot)$ would be close enough to the manifold of physically meaningful states so that convergence to a desired state as input to a reaction happens in controllable time. 
One advantage of the Hartree-Fock approach will likely be that this has considerably lower constant factors than the scattering approach.  Among the advantages of the scattering approach are that it should work for general states (including those where Hartree-Fock does not provide a sufficiently good approximation) and that it provides initial states closer to nature. 
}

As mentioned above, the embedding processes into a larger environment play an important role in the framework during the molecular preparation stage. The ability to dissipate excess energy is essential for both the probability of successfully forming stable bound states and the ability of the dynamics to emulate the open-system evolution of chemical experiments. 
\reviseRepl{Recently proposed methods}{The framework, therefore, employs recently proposed methods} for simulating the weak coupling to a large, memory-less (i.e., Markovian) environment\added{ modelled by a Lindbladian~\cite{lindblad1976generators,mehra1972some} can be used to model the presence of such a dissipative bath \cite{cleve2017efficient,chen2023quantum,ding2024lower,pocrnic2023quantum,hagan2025thermodynamic}.}.
This simulation is efficient in the size of the system and for some methods shown to converge to a thermal state \cite{chen2023quantum,hagan2025thermodynamic}. The only parameter of this procedure that does not scale polynomially with system size is the thermalization time, which is difficult to predict and can, in principle, grow exponentially with system size. However, our observation is that slowly thermalizing systems in our simulation correspond to systems that also thermalize slowly in nature. Thus, for physically motivated open-system models, we would expect to produce either thermalized, metastable, or slowly thermalizing states, depending on which ones are prevalent in nature.  A good example of these types of systems would be a glassy molecular mixture. From this perspective, we conjecture that polynomial-time simulations are sufficient to reach all chemically relevant states. 

One thing to note is that the system on which the readout is to be performed may need to include some degrees of freedom surrounding the molecule, e.g., if solvent effects, photon or phonon couplings, or non-Markovian dynamics~\cite{dan2024qheom} are important. The preparation of this more explicit bath follows the same framework as the main molecular degrees of freedom.

\section{A Scattering-based State-preparation Step}\label{sec:scattering-via-mergo-etc}
\reviseNew{
With a conceptual framework in mind, in this section, we discuss a specific instance of a single scattering step. The approach is based on an external-potential assisted merging of molecular fragments and the construction  of a success-heralding measurement oracle.
In the first part, we discuss the coherent part of the simulation of the merging. Part of the open-system character of the approach in \cref{fig:framework} is brought into play by means of the measurement oracle outlined in the second part of this section.
}

\subsection{Molecular States by Mergo-Association}\label{sec:merging-mols-by-mergoass}
\reviseNew{
A promising approach to realize the assemblage of molecular wavefunctions in \cref{fig:framework} is mergo-association, as considered in \cite{ruttley2023formation,bird2023making,bird2024makingmoleculesmergoassociationrole}. 
Here, optical tweezers confine two fragments -- e.g., two Hydrogen atoms -- to form a bond.

To mimic this, we propose to carry out the simulation in the following way. We consider a real-space first-quantized representation the most convenient with oracles for the repeat-until-success procedure.
We closely follow the implementation put forward in \cite{su_fault-tolerant_2021} to represent the Hamiltonian.
However, note that the core building block is open-system simulation, e.g., implemented by combining the methods of \cite{su_fault-tolerant_2021} with open-system techniques, as proposed in \cite{chen2023quantum}.
Then, we place the two participating systems at a reasonable estimate for a bonding distance. For Dihydrogen, this is well-known; for more involved systems, a heuristic guess needs to be found. 
Initially, we implement their dynamics according to two independent Hamiltonians, $H_A$ on subsystem $A$ and $H_B$ on subsystem $B$. 
Then, we slowly turn on inter-system Coulomb interactions $H_{AB}=V^{\rm int}$ together with trapping potentials $V^{\rm trap}$, according to
\begin{equation}\label{eq:scheduling-the-traps0}
    H(s) = H_A + H_B + f(s)H_{AB} + g(s) V^{\rm trap}.
\end{equation}
For the example of two nuclei, modeling the trap by a harmonic potential acting on nuclear coordinates $R_1, R_2$, its functional form is given simply by~\cite{ruttley2023formation,bird2023making,bird2024makingmoleculesmergoassociationrole}
\begin{align}
    V^{\rm trap}(R_1, R_2) &=  V^{\rm trap}_1 (R_1) + V^{\rm trap}_2 (R_2), \\
    V^{\rm trap}_j(R_j) &=  \frac{m_j}{2} (R_j - R_{0,j})^T \omega_j^2 (R_j - R_{0,j}), \quad j=1,2. \label{eq:trap-per-coo}
\end{align}
More details on this are in \cref{app:mergoass-encoding}. 
A qualitative choice of $f, g$ is displayed in \cref{fig:trap-int-scheduler}. 
After the merging ($s\ge s_0$), the trap is re-released ($s_0\le s \le s_1$), while the interaction stays on. With a certain probability of success, the state will not undergo a diabatic transition into a higher-lying excited state, meaning a successful merger has occurred. 
The specific design of scheduling functions is left up to further research. However, in order to mimic mergo-association, we propose the use of an $f(s)$ that schedules the inter-species interactions that resembles the trajectory of a Coulomb potential. 
More details on a quantum encoding of the trap potential can be found in \cref{app:mergoass-scattering,app:mergoass-encoding}.

Equivalently to the adiabatic evolution in $s$, one may put system $A$ and system $B$ at initial distance $z^*$ and move their centers of masses together at a predetermined rate $\frac{{\rm d} \Delta z}{{\rm d}t}$ until the for bonding length is reached. Hence, the scheduling functions $f(s), g(s)$ are not present in the Hamiltonian anymore, and naturally the strength of the interactions is implicitly determined by distance -- recovering the procedure of mergo-association. 

The mergo-association scheme we follow relies on assumptions from scattering theory. 
The setup in scattering theory is that two scattered objects, situated beyond a certain scattering length, can be modeled as free particles. Within the scattering length, the scattered product is considered as `one'. 
Conceptually, this is analogous to the `molecule factory' in \cref{fig:framework}.
The scattering length, or harmonic length, is defined as $\beta = \sqrt{\hbar/(\mu\omega)}$ with a frequency $\omega$ and relative mass $\mu = m_1m_2/(m_1+m_2)$ and induces a notion of a bonding energy via $\hbar\omega$. 
As we are interested in chemical bonds, a proxy for this scattering frequency can be the most weakly bound state or the highest-lying \textit{bonded} vibrational excited state.

\begin{figure}[h]
    \centering
    \includegraphics[width=0.75\textwidth]{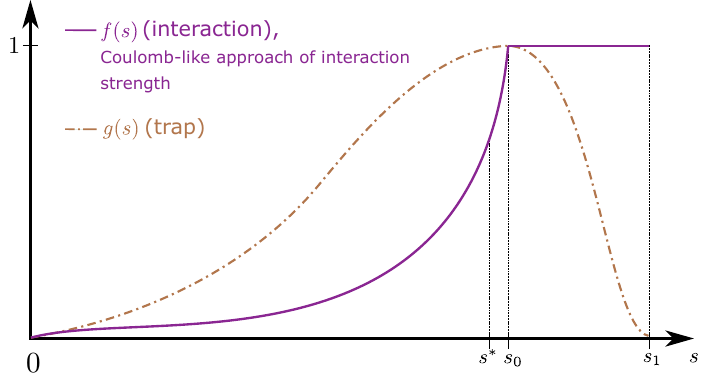}
    \caption{Scheduling of interactions. Inter-species interaction is described by $f(s)$ so that $f(0)=0, f(s\ge s_0)=1$ and is monotonically increasing until $s_0$ and then constant. The harmonic trap follows $g(s)$ with $g(0)=0, g(s_0)=1, g(s_1)=0$. It is monotonically increasing until $s_0$ and then decreasing until $s_1$. Following adiabatic evolution, there is a `point of contact' of states at $s^*$ a little earlier than $s_0$, which is the evolution parameter (with corresponding effective distance) used to evaluate the diabatic transition probability. Here, we assume the scenario that the constituents are already placed at closed distance and we `slowly turn on' the Coulomb interaction, where with the trajectory of $f(s)$, we aim to mimick an evolution of interaction strength that is Coulomb-like if they were to approach each other. 
    To that end, let $z_{0,1},z_{0,2}$ be the centers of the traps. Then, in order to follow $V^{\rm int}(R_1(s), R_2(s)) = \frac{q_1 q_2}{\left|  \Delta R_{12}(s)\right|}$ with nuclear charge $q_j$, let the implemented interaction be $ V^{\rm int}(s)=f(s)H_{AB} =  f(s) \frac{q_1 q_2}{\left| z_{0,1}-z_{0,2}\right|}$ where via $f(s)= \frac{|z_{0,1}-z_{0,2}|}{|\Delta R_{12}(s)|}$, the desired motion of the nuclei can be emulated. Towards $s \to s_0$, the fixed-strength potential assuming trap centers needs to be replaced by the actual strengths to account for fluctuations in the positions which will matter at that stage.}
    \label{fig:trap-int-scheduler}
\end{figure}

The key aspect of such a mergo-association is the choice of trap frequency $\omega$ and the consequences on the diabatic transition probability. Here, we work with an isotropic trap for simplicity, hence \cref{eq:trap-per-coo} is described by a single trap frequency $\omega$.
Successful, in our case, means that a (e.g., covalent) bond has been formed and that there is no diabatic transition beyond a certain vibrational frequency of the \textit{trapped} system, say $\omega_a$. 
Then, it is possible to approximate the probability of success by using the Landau-Zener probability~\cite{wittig2005landau}, 
\begin{equation}
    p_{\rm suc} \ge 1 - {p}_{\rm LZ}; \quad\quad {p}_{\rm LZ} \approx
    \exp( -2\pi \Gamma ) \quad {\text{   with   }} \quad  \Gamma = \frac{\omega_{\rm eff}^2 / \hbar}{ \left|\frac{\rm d }{{\rm d}s} (E_{\rm mol} - E_{\rm atom})\right|}
\end{equation}
Here, $\omega_{\rm eff}^2$ is the effective potential strength of the trapped system (see \cref{app:psuc-discussion}).
Then, 
we use that the scheduling function $f(s)$ can be related to a corresponding inter-nuclear motion $R(s)$ in a  physical mergo-association experiment. Thus, we can rewrite the expression in terms of derivatives of this equivalent position, $\frac{\mathrm{d} E}{\mathrm{d}s} = \frac{\mathrm{d} E}{\mathrm{d}R} \frac{\mathrm{d} R}{\mathrm{d}s}$. More details are in the caption of \cref{fig:trap-int-scheduler}.
Defining the velocity $v = \frac{{\rm d}R}{{\rm d}s}$ and using the shorthand $\partial E_{\rm mol}, \partial E_{\rm atom}$ for the energy derivatives, this gives 
\begin{equation}
    {p}_{\rm LZ} \approx
    \exp( -\frac{2\pi}{\hbar}\frac{\omega_{\rm eff}^2}{\lvert \partial E_{\rm mol} - \partial E_{\rm atom} \rvert \, v}   )
\end{equation}
A more detailed discussion can be found in \cref{app:psuc-discussion}, and we continue by presenting a simplified final expression. 
We start by the observation in \cite[Eq.~(54)]{bird2023making} that 
\begin{equation}\label{eq:oeff-via-rel-bond-en}
    \omega_{\rm eff}^2 \appropto   \omega^2 \tilde E_a^{1/2}\exp(-\tilde E_a)
\end{equation}
with $\tilde E_a = \frac{\hbar\omega_a}{\hbar\omega}$ as a `relative bonding energy'. 
Together with expressions for energy derivatives from \cref{eq:expression-energy-derivative} and an approximation to \cref{eq:oeff-via-rel-bond-en} we can obtain a bound on $p_{\rm LZ}$,
\begin{equation}\label{eq:bound-on-landau-zehner}
    p_{\rm LZ} \lessapprox
    \exp(
    -4\left(\frac{\pi}{\mu}\right)^{1/2}
    \left( 
    \hbar\omega
    \frac{
        \tilde E_a
    }{
    3 + \tilde E_a
    }
    \right)^{1/2}
    \frac{
    \exp(-\frac{1}{2}\tilde E_a - \frac{3}{2})
    }{v}
    )
\end{equation}
Ideally, we want $p_{\rm LZ}$ to be close to zero so that $p_{\rm suc}$ is close to one. 
Then, up to the speed of evolution $v$ and effective mass $\mu$, this depends on the ratio of binding energy compared to trap energy, $\tilde E_a \sim \omega_a/\omega$. 
The inner exponential is closer to one (which means smaller $p_{\rm LZ}$) for small $\tilde E_a$ and thus for $\omega$ as large as possible relative to $\omega_a$. 
Beyond this, for fixed $\hbar\omega_a$, the other term depending on frequencies, $\left( 
    \hbar\omega
    \frac{
        \tilde E_a
    }{ 3 + \tilde E_a
    }
    \right)$, is also maximal for large trap strength $\omega$.
However, there is a trade-off in maximizing this probability of success versus minimizing the cost factor of the block-encoding of the operator. The latter represents the cost of representing the potential in the block-encoding access model, and is quantified by the operator one-norm of the potential, $\alpha_{\rm trap}$. Generally, we would aim to keep the frequency no larger than necessary to keep the block-encoding cost low.
Additionally, we can look at the system of interest in \cite{ruttley2023formation,bird2023making,bird2024makingmoleculesmergoassociationrole}, namely Rb and Cs.
They use $\omega=$150~kHz for all of their simulations and experiments.
This order of magnitude is about the same as the appearing $\omega_a$, which is roughly $110\,{\rm kHz}$; the scattering length (regarding binding energy) for this system \textit{in a weakly bonded regime}, is approximately $645\,a_0$~\cite{takekoshi2012towards}.
This setup allows \cite{ruttley2023formation} to observe high probabilities of success (see \cite[Fig.~4]{ruttley2023formation}, close to 80\,\%). 

Typical covalent bonds are of the order of $100\,\frac{\rm kcal}{\rm mol}$, which amounts to approximately a $35,000\,\frac{1}{\rm cm}$ frequency for $\omega_a$ (approx. $10^{12}\,$kHz). Therefore, we expect the required  $\omega$ to be, in general, vastly larger than $150\,{\rm kHz}$; however, the hope is that the remaining behavior will carry over, and we only need to scale it according to the relative difference in bonding energy. Then, in fact, $E_a=\hbar \omega_a$ can remain bounded in the two-atom/molecule scatterings and effectively act as a constant in an asymptotic sense.
This means for $p_{\rm LZ}$ to be bounded as well, $\omega$ incurs a (likely substantial) constant factor through $E_a$ and then scales as $O(\mu)$ for $\mu=\frac{m_1m_2}{m_1+m_2}$. Therefore, since at the last scattering stage in the ``molecule factory'', $\mu\sim \numnuc$ , we can roughly estimate $\omega$ in the expression of $\alpha_{\rm trap}$ by a linear factor $\numnuc$ and a system-dependent constant factor for the binding energy of covalent bonds. 
Using this in our asymptotic expression for the block-encoding factor of the trap, 
\begin{equation}\label{eq:trapped-adiab-hamiltonian}
    \alpha_{\rm trap} \approx O(\numnuc^3 \Omega_{\rm trap}^{2/3}).
\end{equation}
We want to compare this result with the corresponding sub-normalization factors of the kinetic energy, $\sim \frac{\numel N^{2/3}}{\Omega^{2/3}}$, where $N$ is the number of particles and $\Omega$ the volume of the simulation box, and the Coulomb energies, $\sim \frac{\numel^2 N^{1/3}}{\Omega^{1/3}}$~\cite{su_fault-tolerant_2021}.
If the number of grid points grows with the number of particles, then the encodings of these operators increase roughly with power $\frac{5}{3}$ or $\frac{7}{3}$, whereas the trap scales cubically. It also grows with volume $\Omega^{2/3}$ rather than with the inverse grid density. 
A more accurate representation will need a finer grid spacing, increasing the cost of representation for the Coulomb potentials. For the trap potential, the box size matters, which will need to increase for larger systems, although there is no increase in cost for increasing accuracy through a finer grid spacing.
} 

\subsection{Measurement Oracles for Success Heralding}\label{subsec:meas-oracles-main-text}
\reviseNew{
Suppose now that after a scattering step, we aim to herald success according to \cref{eq:success-channel,eq:non-success-channel}. To that end, we need a measurement operator $\mathbb O_{\rm suc}$ able to give the desired partitioning. 
A more detailed discussion of the following outline can be found in \cref{app:meas-oracles-appendix}.
} 

\subsubsection{Success of merging}
\reviseNew{
First, we start by briefly outlining the success heralding via weak measurements. 
We are given a quantum state that results from a previous `scattering' event, such as by simulating the Hamiltonian in \cref{sec:merging-mols-by-mergoass}, which is supported by a trap potential. 
Such a state will be, following the notion of success that we used before, in a superposition of states that underwent a diabatic transition into higher-lying vibrational states or not. 
This notion of success does not correspond with one that we can directly observe in a non-destructive manner. Thus, we develop another notion of `success' that we can use for efficient success heralding to ensure a successful merging. 
The state we assume is in a first-quantized real-space discretization; this may not be an optimal basis but is most convenient for our discussion, and efficient transformations to other bases can be thought of as well~\cite{babbush_encoding_2018,bagherimehrab2024efficient}. 
Then, success can be formulated most straightforwardly using proximity as a geometrical criterion. E.g., suppose we collect a set of locations of interest where we want to test proximity across the nuclei, $\mathcal{I} \subseteq [\numnuc]\times[\numnuc]$, then 
\begin{align}
    \tilde{\mathcal{C}}_{\text{geom}, \Delta} (\{r,R\}, \mathcal I)  = \begin{cases} 1 & \left\|R_j - R_k\right\| \leq \Delta_{jk} \quad \forall (j,k)\in \mathcal{I}\\
    0 & \text{else.}
    \end{cases}\label{eq:classical-criterion-main-text}
\end{align}
with $\Delta_{jk} >0$. We mark success if the criterion $\mathcal{C}$ returns 1. Note that we choose to restrict the evaluation to nuclear coordinates as the relative position of nuclei within a bonded state is essentially stationary compared to the electrons.
The definition in \cref{eq:classical-criterion-main-text} is classical and assumes direct access to $\{R_j\}_{j\in[\numnuc]}$, whereas types of states we encode are superpositions of tensor products across basis state labels, $\ket{\mathtt r_1, \mathtt r_2,\cdots, \mathtt r_{\numel};\,\mathtt R_1,\cdots,\mathtt R_{\numnuc}}$. 
More generally, we can think of what happens to a mixed state when going through the success heralding. 
Then, we consider states of the form
\begin{align}
    \rho = \sum_{p,q} \rho_{p,q} \op{\psi(\{r,R\}_p)}{\psi(\{r,R\}_{q})}, \quad \small{ \ket{\psi(\{r,R\}_p)}\sim \big\vert \mathtt r_1^{(p)}, \mathtt r_2^{(p)},\cdots, \mathtt r_{\numel}^{(p)};\,\mathtt R_1^{(p)},\cdots,\mathtt R_{\numnuc}^{(p)}\big\rangle}.
    \label{eq:general-mixed-state-merging-success}
\end{align}
The set $\{r,R\} = \{r_j\}_{j=1}^{\numel}\cup \{R_j\}_{j=1}^{\numnuc}$ denotes a nuclear and electronic configuration, and $\{r,R\}_j$ denotes an instance as the $p$th term of the superposition. Configurations include a spin degree of freedom. Hence the range of $p,q$ is the number of possible basis states labeling the grid points.

Now, we assume that there is a quantum circuit that can achieve the action of $\mathcal C$ efficiently, namely a $U_{\mathcal C}$ that takes the state $\rho$ plus an ancilla and stores success in the ancilla, according to the following sets:
\begin{align}
    A &= \{ j : \mathcal{C}(\{r,R\}_j)=1 \} \nonumber\\
    B &= \{ j : \mathcal{C}(\{r,R\}_j)=0 \}.
\end{align}
The crucial part here is that such a criterion induces a bi-partition of the Hilbert space, splitting the set of states into two groups. 
Specifically, every state of the form $\ket{\psi(\{r,R\}_j)}$ either corresponds to a configuration with $\mathcal{C}(\{r,R\}_j)=1$ or a configuration with $\mathcal{C}(\{r,R\}_j)=0$. 
The set $A$ enumerates the states the oracle accepts as merged, and $B$ enumerates the states that the oracle rejects. For convenience we will denote $\ket{\psi_j} :=  \ket{\psi(\{r,R\}_j)}$ from now on. 
With this bi-partition, a general input state can now be written as
\begin{align}
    \rho =& \sum_{j,k \in A} \rho_{j,k} \op{\psi_j}{\psi_k} + \sum_{j,k \in B} \rho_{j,k} \op{\psi_j}{\psi_k} 
    + \sum_{j \in A, k \in B} \left( \rho_{j,k} \op{\psi_j}{\psi_k} + \rho_{k,j} \op{\psi_k}{\psi_j} \right) \label{eq:input_to_oracle}
\end{align}
Weak measurement of this oracle and thus flagging of a successful merging can be implemented following the subsequent operations based on work in  \cite{lund2011efficient,andres2022weakly,yan2022fixed,mizel2009critically}:
\begin{enumerate}\label{enumerate-here}
    \item Append an ancilla qubit, $q_{a,1}$, and perform the unitary $U_{\mathcal{C}}$ on the joint system. This stores the `success value' in the ancilla.
    \item Append a second ancilla, $q_{a,2}$, and perform a controlled rotation CR${}_y(\delta)$ gate on $q_{a,2}$ conditioned on $q_{a,1}$.
    \item Perform $U_{\mathcal{C}}$ on the input state and $q_{a,1}$, resetting $q_{a,1}$.
    \item Measure $q_{a,2}$ in the computational basis and reset it for later use. The measurement on $q_{a,2}$ then is used as a result flag of a successful merger.
\end{enumerate}
Finally, we can conclude that the measurement oracle $\mathbb{O}_{\rm suc}$ is composed by the circuit $U_{\mathcal C}$ and the weak measurement scheme above.
More details are given in \cref{subsec:oracle_on_superpositions}.

An important aspect to consider in constructing these measurements is that the wavefunctions encoded, even when expressed as a mixed state, need to satisfy the necessary exchange symmetry (antisymmetric for Fermions, symmetric for Bosons). 
Ongoing measurement induces the risk of compromising such symmetries. Thus, the measurement needs to be adapted to satisfy that. A detailed discussion is provided in \cref{subsec:distinguishable-oracles}.
} 

\subsubsection{Spin}\label{subsubsec:spin-oracle}
\reviseNew{
It is straightforward to append $\mathbb{O}_{\rm suc}$ by a check whether the obtained state is in the correct spin state. Suppose we have access to an implementation of the time-evolution of the (normalized) spin operator $S^2$. 
Then, one step of quantum phase estimation allows storing in an ancilla qubit whether there is a singlet~(0) or triplet~(1) state. Augmentation to higher-order spin states (such as doublets, quadruplets, etc.) follows simply by augmenting the ancilla register of the QPE circuit to represent the appropriate spin numbers.  
Based on this result, we can accept or reject based on the obtained result and thus effectively have a scheme that projects into the desired spin state. 
Therefore, this yields an implementation of a projection into the correct spin state; this requires nontrivial knowledge about the expected spin state. 
One can draw from ideas in \cite{jackson2022optimal} to deal with more complicated ensembles of spin states.
A detailed procedure in case of lack of knowledge or intuition about the expected spin state is subject to further research.

We remark that construction of the $S^2$ operator is simple given an encoding as in \cref{eq:general-mixed-state-merging-success}, as then it simply translates to its form as Pauli operators acting on the spin degree of freedom.
} 

\section{Measuring Dynamical Quantities of Interest and a Review of Exemplary Applications}
\begin{figure}
    \centering
    \includegraphics[width=.6\linewidth]{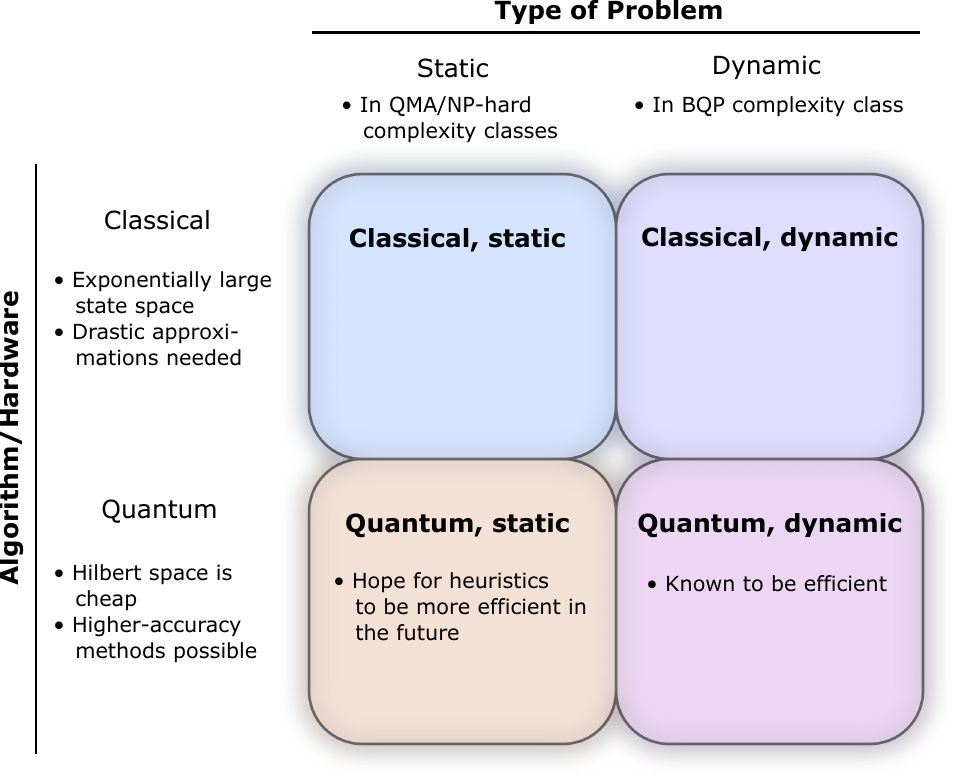}
    \caption{Classifying chemical problems related to their hardness and space complexity. Dynamical properties are quantumly efficient, whereas static properties are generally hard. As quantum computers do not suffer from the curse of dimensionality, one can expect the sweet spot of quantum simulations, up to constant factors in the cost, to lie in the evaluation of dynamic properties.  }
    \label{fig:problem-classes}
\end{figure}

Simulating the dynamics of molecular processes provides access to meaningful information about the rearrangements of atomic nuclei and electrons and their interactions with electromagnetic fields as they unfold. Already with classical resources, time-dependent approaches offer numerous advantages over time-independent ones, as the former are more amenable to handling the continuum portion of the spectrum and grant access to the relevant elements of the scattering matrix \reviseRepl{\footnote{The $S$-matrix has transition elements concerning a process represented by the \textit{evolution} step in \cref{fig:framework}, independent of the state preparation scheme in the \textit{molecule factory} part.}}{($S$-matrix)}  over a meaningful range of energies. We provide an overview of chemical problems, with time-dependent versus stationary quantities, solved on different hardware, in \cref{fig:problem-classes}. 

Measuring observables in a dynamical picture requires considering the time evolution of a wave packet, i.e., a superposition of solutions of the time-dependent Schr\"odinger equation (TDSE). Most measurements of dynamical quantities can be phrased in terms of wave packet correlation functions, whose calculation fits perfectly into our framework. Transition amplitudes are measured, e.g., using a Hadamard test~\cite{kassal_polynomial-time_2008,pedernales_efficient_2014,kokcu_linear_2023}.  We can follow the scheme introduced in~\cite{tacchino_quantum_2020}, which is capable of obtaining two-point correlation functions, or the cumulative correlation function method in~\cite{lin_heisenberg-limited_2022}. This scheme is extendable to the S-matrix and to $n$-point correlators by additional time-evolutions and block encodings, similarly to the linear response framework in~\cite{kokcu_linear_2023}. If results at more than two times are required, we can use a history state encoding following the conditional time evolutions of the quantum circuit of quantum phase estimation, similar to the construction in \cite{diaz2023parallel}, who also consider observables measured across various time-steps. To that end, we add a clock register $\sum_{\ell=0}^{n_t} \ket{\ell \Delta t}$  so that $n_t \Delta t=t$. Then, instead of a direct time evolution of the overall system after the molecule factory in \cref{fig:framework} in the state $\rho_1^\text{mol}\rho_2^\text{mol}\cdots\rho_M^\text{mol}$, we split the evolution into chunks of $\Delta t$ and condition on the clock register to produce a superposition of the state at different evolution time steps. Furthermore, if needed, applying a Quantum Fourier Transform before measurement is straightforward and does not compromise the overall efficiency of the algorithm.  
While this framework could be interpreted as purely theoretical, we can easily show that this approach covers a vast range of contexts, extending from nature to chemical laboratories. 

\paragraph*{\textbf{Reaction rates.}}
Chemists are commonly interested in observed kinetics, i.e., reaction rates, a topic inherently suitable for quantum dynamics simulation. For instance, dynamical simulation for nonadiabatic processes such as charge transfer reactions falls into the class of quantum circuits of polynomial complexity~\cite{ollitrault_nonadiabatic_2020}. Our framework allows using the measurement schemes proposed in~\cite{kassal_polynomial-time_2008} in which the reaction rate can be computed from the degree of localization of the wave packet in nuclear configuration space corresponding to the formed products. In addition to making the simulation itself achievable, the quantum approach reduces the measurement to a binary search instead of the more complicated evaluation of a time correlation function~\cite{liu_prospects_2022}. \reviseRepl{If needed, the latter can also be computed from the dynamics via a rate constant calculated as a function of the scattering cross-sections.}{Nonetheless, if needed, the latter could, in principle, be computed from the dynamics so that the rate constant is calculated as a function of the scattering cross-sections.}

\paragraph*{\textbf{Photochemistry and photophysics.}}
Photochemical processes are triggered when a molecule absorbs a photon. In these transformations, electronically excited states become populated, giving access to reactive channels that are thermally unachievable~\cite{bhuyan2023rise}. The interaction between a molecule and an external photonic field alters the potential energy surface on which wave packets propagate. Therefore, the nuclear and electronic degrees of freedom cannot be decoupled, and the Born-Oppenheimer approximation breaks down~\cite{curchod_ab_2018}. The dynamics simulation of these systems on classical computers becomes prohibitive after including very few degrees of freedom. 
The inclusion of photonic fields in the procedure according to \cref{fig:framework} is straightforward by explicitly adding a register for photonic degrees, an explicit bath, or a modified Hamiltonian as in \cref{eq:trapped-adiab-hamiltonian}.
\reviseRepl{Further photophysical processes that can be simulated are, for instance, when spectroscopic measurements do not suffice, ultrafast spectroscopy, which tracks the system during a light-initiated transformation. Other examples are light-harvesting complexes, molecular machines, and photovoltaics; recent work in \cite{motlagh2024quantum} also proposed a dynamics-based algorithm for singlet-fission solar cell design. }{Other interesting molecular photophysical processes are also within the scope of our framework. For instance, spectroscopic measurements rarely go to orders higher than second, and instead, ultrafast spectroscopy is preferred, where the system is tracked during a transformation initiated by light. Light-harvesting complexes, photovoltaics, and molecular machines are examples of photophysical applications that our framework can simulate.} Our approach would make it possible for quantum computers to simulate quantum systems, e.g., optoelectronic devices, where the quantum dynamics are far more important than the eigenstates of the Hamiltonian. In these devices, the relaxation pathways of the excitons are exploited for light generation and harvesting. Classical simulations suffer from the large space of excitons and phonon coupling, making current simulations beyond hopeless. Conversely, simulating the time evolution of thermal quantum states inherently captures all the necessary behavior.

\paragraph*{\textbf{Linear and non-linear molecular spectroscopy.}}
\reviseRepl{Another class of problems that dynamics can solve is the }{
Our proposed framework enables computations connected with the} in-laboratory characterization of molecular structures and properties. The absorption spectrum of a molecule is given by the Fourier transform of the wave packet autocorrelation function~\cite{tannor_introduction_2007}. By judiciously selecting an initial state, measuring the autocorrelation function can provide us with different spectra, including electronic, vibrational, and rotational~\cite{vitale_anharmonic_2015}. Through the computation of emission and absorption spectra, our approach accommodates the simulation of fluorescent systems, such as those used in biomolecule marking or thermally activated delayed fluorescence, where forbidden relaxations are assisted by thermal coupling to the environment. Additionally, $n$-time correlation functions allow the exploration of linear-response molecular spectroscopy beyond UV-Vis and fluorescence, such as rotational or vibrational spectroscopy, in which contributions from excited states are typically significant at room temperature. Simulating spectroscopic measurements can be used not only to reproduce experiments but also to probe the simulated quantum system, e.g., the presence of an IR signal may indicate the formation of a molecule, as in the molecular factory in \cref{fig:framework}. 
Two-dimensional spectroscopy could also benefit enormously from extracting $n$-time correlation functions from dynamics by following our approach. For instance, two-dimensional infrared (2D IR) spectroscopy reveals second-order vibrational couplings, which characterize molecular interactions~\cite{thunt_2d-ir_2009,noda_two-dimensional_1990}. Classical simulations of this state-of-the-art technique typically exclude anharmonicity and produce errors associated with the BO approximation and vibrational population transfer. In this example and many others, finding the ground state of the system is far removed from reproducing the experimental spectrum.

\paragraph*{\textbf{Free Energy Simulations.}}
Free energies play a role in many naturally occurring physical processes\reviseRepl{, as they determine }{. The free energy determines} whether a process occurs spontaneously, 
such as whether a ligand binds to a protein, whether a material such as salt dissolves in water, or into what shape a protein folds. The type of free energy relevant to a specific system depends on the nature of the system. In the case of an isothermal, isobaric system that only allows volumetric work, the relevant free energy is the Gibbs free energy. In applications, one is primarily interested in free-energy differences between two different states of the system, which can be characterized by two Hamiltonians, $H_1$   and $H_2$. There exist multiple methods for calculating the free-energy difference between these two states. One class of methods uses fluctuation relations, such as the Jarzinsky equation, that require ensembles of dynamics simulation of the systems of interest~\cite{bassman_oftelie_computing_2022}. 
To estimate the free-energy difference, these methods require evaluating the total energy of each trajectory for both the initial and final states. A time-dependent Hamiltonian $H(\lambda(t))$ is used to alter the system from state 1 to state 2, where $\lambda$ is an externally controlled parameter such that  $H(0) = H_1$   and  $H(1)=H_2$   (the transition is generally non-adiabatic, and states do not have to be or remain in the ground state such as in adiabatic quantum computing). The number of energy values required depends on the type of problem considered, the speed at which the Hamiltonian is transformed, and the desired accuracy. There are two straightforward ways to calculate free energies in alignment with \cref{fig:framework}. First, one can create a sufficiently large number of identical systems and perform separate quantum simulations before measuring the total energies of the initial and final states. 
Alternatively, one can use the clock-register approach from above and simulate a single trajectory.  To obtain a sufficiently large number of statistically independent energy values of both initial and final states, one can simulate each state long enough to take measurements with sufficiently large time intervals in between. Once  \reviseRepl{individual energies are measured}{the energies are obtained}, the averages and final calculation of the free energy difference can be carried out on a classical computer. 
\added{This approach is expected to be preferred in terms of quantum and classical resources compared to representing a complete ensemble as long as the variance of single-trajectory estimates is not significantly larger than the joint estimate.}

\paragraph*{\textbf{Quantum Machine Learning.}}
Instead of directly measuring interesting quantities, \deleted{it can be envisioned that} the output of the dynamical quantum simulation \reviseRepl{can be processed by quantum machine learning}{is processed using machine learning}. This framework can be envisioned as machine learning with quantum input data, with possible classical or quantum outputs~\cite{biamonte_quantum_2017,cerezo_challenges_2022}. \reviseRepl{In particular, recent results indicate}{This is particularly promising in light of recent results indicating} that there is a provable advantage in the efficiency of extracting information when given access to multiple copies of a state in a format that a quantum computer can manipulate compared to having access to only measurements performed on the state~\cite{huang_quantum_2022}. Thus, certain chemically relevant properties of states and quantum evolutions may be learnable more efficiently in our quantum computing framework than in a conventional experiment. Alternatively, it is also possible to measure spectroscopic quantities and replace physical experiments in machine learning pipelines that operate on spectroscopic results, potentially aiding further developments in molecular design~\cite{angulo_machine_2022,joung_deep_2021}.       

\section{Conclusion and Outlook}
We have provided an algorithmic framework that, in principle, can solve a broad set of chemically relevant problems using inherently efficient building blocks. To that end, we considered that, while general ground states are hard to obtain, we may assume to be able to prepare atomic ones as a single-cost effort that can enable a building block library. A scattering process, implemented by simulating dynamics and boosted by artificial potentials, can produce a molecular input state for a subsequent dynamical simulation, which is then followed by the measurement of dynamical quantities. 
\added{Preparing molecular states via mergo-association is a particularly promising candidate here, which we discussed in detail.}
We provide examples of applications from, e.g., spectroscopy, photochemistry, and beyond. 

Future work remains to build upon this approach and provide \reviseRepl{ a more general and detailed analysis of mergo-association and numerical experiments of this approach}{ a more concrete algorithm}, e.g., to gauge the feasibility of the procedure with respect to constant factors and to investigate the more precise costs arising from choosing specific problem instances and methodologies\deleted{, for example, the detailed computational cost of using artificial potentials to boost success probabilities of the scattering process}. 
Another interesting extension would be the inclusion of other modeling tools used in classical simulation, such as the Nosé-Hoover thermostat~\cite{william_graham_hoover_computational_1991}. \added{Additionally, an interesting avenue would be to consider additional classical dynamical simulation such as more molecular dynamics (e.g., studied in the context of protein modeling in \cite{liu2024toward}), in particular in the case when conservation laws (such as energy conservation) allow direct mappings to Hamiltonian simulation~\cite{babbush2023exponential}.}

A common focal point in quantum chemistry is finding ground state energies, a problem known to be QMA-complete for local Hamiltonians like those seen in molecular systems. We should not restrict ourselves to that perspective, which focuses on a problem known to be hard, and look for paths that allow using dynamics more directly. Additionally, for situations when the ground state is of interest, it may prove useful to give up the search for exact solutions to a hard problem and look at heuristics. The Hartree-Fock problem is known to be NP-hard yet practically efficient, thanks to approximations~\cite{whitfield_computational_2012}. Attempts based on open systems dynamics such as in~\cite{ding2024single,li2024dissipative} may be a promising path towards quantum heuristics for ground states. Nevertheless, we call upon what dynamical quantum simulation offers for chemistry.

\section*{Acknowledgement}
The authors thank Juan Bernardo Perez Sanchez, Kouhei Nakaji, Mohammad Ghazi Vakili, FuTe Wong, Sumner Alperin-Lea for helpful discussions and comments. We further thank Jakob~S. Kottmann for confirming that LBK is always right.

AA gratefully acknowledges King Abdullah University of Science and Technology (KAUST) for the KAUST Ibn Rushd Postdoctoral Fellowship. LBK acknowledges support from the Carlsberg Foundation.
This research was developed with funding from the Defence Advanced Research Projects Agency (DARPA) and the Department of the Navy, Office of Naval Research. The views, opinions, and/or findings expressed are those of the authors and should not be interpreted as representing the official views or policies of the Department of Defense or the U.S. Government. AAG acknowledges support from the NSERC Industrial Research Chairs Program and the Canada 150 Research Chairs.
This work is supported by the Novo Nordisk Foundation, Grant number NNF22SA0081175, NNF Quantum Computing Programme.
Resources used in preparing this research were provided, in part, by the Province of Ontario, the Government of Canada through CIFAR, and companies sponsoring the Vector Institute~\url{https://vectorinstitute.ai/partnerships/current-partners/}.

\bibliography{arxiv-update-at-resubmission/chempolybib}

\appendix

\section{Scattering Molecules via Mergo-Association}\label{app:mergoass-scattering}
This section includes more details on merging molecules by mergo-association as described in \cref{sec:merging-mols-by-mergoass}.

\subsection{Outline of Quantum Computational Encoding}\label{app:mergoass-encoding}
First, we describe the quantum computational encoding of the approach.
A grid $G$ of $N$ points labels an integer lattice, where for convenience, we assume that $N$ has an integer cubic root so that $\exists  m \in \mathbb{N},\, m^3 = N$,
\begin{equation}\label{eq:grid-labelling}
    G = \left[-\frac{N^{1/3}-1}{2}, \frac{N^{1/3}-1}{2}\right]^3.
\end{equation}
The associated volume box $\Omega := [-\frac{L}{2}, \frac L 2]^3$. Exemplarity for Dihyodrogen with $\numel=2$, $\numnuc=2$, an estimated volume of $\lvert\Omega\rvert = L^3 = (10 a_0)^3$ is expected to suffice, where $a_0$ is the Bohr radius. With a grid spacing of $\sim 0.1~a_0$, this volume would lead to $N\sim 10^6$.
Then, upon defining the size of the simulation box, it is straightforward to map every $p\in G$ to a coordinate, $\mathtt r_p\in \Omega$ for electrons and $\mathtt R_p\in\Omega$ for nuclei.

The following is needed to describe the desired dynamics: Hamiltonians describing the isolated Hydrogen atoms, interactions between the Hydrogen atoms, and a `trap potential' modeled by a harmonic oscillator potential.
The system Hamiltonian without trap is given by
\begin{equation}
    H_{\rm sys} = T_{\rm el} + T_{\rm nuc} + U_{\rm nuc-el} + V_{\rm ee} + V_{\rm nn}.
\end{equation}
As the initial state to the simulation, we assume access to a product state of the ground states of the Hydrogen atom, respectively,
\begin{equation}
    \lvert \psi_{0,\rm H} \rangle  \lvert \psi_{0,\rm H} \rangle.
\end{equation}
From \cite{su_fault-tolerant_2021}, we know the sub-normalization factors (which we will repeat further below) to encode these in real space, together with estimates for the cost of Hamiltonian simulation algorithms. 
The placement of the individual $\ket{\psi_{0,\rm H}}$ should follow a guess for the desired molecular distance so that the relaxation through the trap potential induces a more accurate placement for the bonded state. 

The harmonic trap potential from \cite{bird2023making,bird2024makingmoleculesmergoassociationrole} is
\begin{align}
    V^{\rm trap}(R_1, R_2) =  V^{\rm trap}_1 (R_1) + V^{\rm trap}_2 (R_2),
\end{align}
which confines the nuclear motion per atom $j$:
\begin{equation}\label{eq:v-rel-trap-vectors}
    V^{\rm trap}_j(R_j) =  \frac{m_j}{2} (R_j - R_{0,j})^T \omega_j^2 (R_j - R_{0,j}).
\end{equation}
We may neglect electronic mass and coordinates for the trap parameters given the large discrepancy between electronic and nuclear mass; the confinement is a heuristic approach nonetheless. 
The positions $R_{0,j}$ define the center of the two traps. For single-particle nuclei, such as Dihydrogen, the determination of trap centers is straightforward; in contrast, for larger systems, center-of-mass and relative coordinates need to be considered~\cite{bird2024makingmoleculesmergoassociationrole}.
The frequencies $\omega_{j}^2$, in general, make up positive, diagonal matrices; the expression above describes the strength of the trap, which may not be spherically symmetric. 
For the sake of our discussion here, we will assume the trap to be isotropic, and then the trap frequency can be described by a scalar $\omega_j$.
The role of these frequencies and relevant energy scales will be discussed further below in \cref{app:psuc-discussion}.

Next, we discuss a quantum implementation of the trap potential. We reformulate \cref{eq:v-rel-trap-vectors} to
\begin{equation}
\begin{split}
     V^{\rm trap}&(R_1,\ldots,R_{\numnuc}) =\\
     =&
      \sum_{j \in \text{nuclei}} \frac{m_j}{2} \sum_{p\in G} \sum_{w\in\{x,y,z\}}\omega^2_{j,w} (R_{p_j,w} - R_{0;w})^2  \op{p}_j,
     \end{split}
\end{equation}
acting on position labeled by $p$.
The state we encode is a superposition over $\ket{p}=\ket{p_1}\cdots \ket{p_{\numel+\numnuc}}$, which hold grid labels for each particle in a register, plus one extra qubit per particle for spin if desired. 
Synonymeously in this work, we use typewriter-font to denote the grid labels, $\ket{\mathtt{r}_1}\cdots\ket{\mathtt r_{\numel}}\ket{\mathtt R_1}\cdots\ket{\mathtt R_{\numnuc}}$. A state called $\ket{r_j}$ or $\ket{R_k}$ will hold the explicit coordinate information.
This requires $\log(N)$ qubits per particle to represent its position on the grid.
For Dihydrogen, overall $\numnuc=\numel=2$, so we need $4\times\log_2(\text{number of grid-points})$ qubits to encode the grid; with the estimated $10^6$ grid points from above, that makes roughly 80~qubits for the $\ket{p}$ register.

The procedure that is simulated is described using scheduling functions as mentioned above in \cref{eq:scheduling-the-traps0}, so that $f(s)$ introduces the inter-atomic Coulomb interactions and $g(s)$ guides the strength of the trap. As a consequence, implementing this scattering follows evolution across 
   $ \mathcal{T} {\rm e}^{-{\rm i}\int_0^{s_1} H(s) {\rm d}s}$,
with $H(s)$ from \cref{eq:scheduling-the-traps0}. The truncated Dyson series algorithm in \cite{low2018hamiltonian} can serve as a way to implement this.
A necessary requirement is then to have access to the potentials as a HAM-T oracle, which we outline for $V^{\rm trap}$:
\begin{equation}
  \sum_{s = 0}^{n_t} \frac{ V^{\rm trap}(s\frac{s_{1}}{n_t}) \otimes \op{s}  }{\alpha_{\rm trap}}.
\end{equation}
Using this, we can simulate over $s$ across the scheduling functions so that the inter-species interactions remain and the trap has fully decayed, i.e., until $s_1$. This operation can be constructed easily given the LCU outlined in \cref{eq:prep-for-trap,eq:sel-for-trap} together with oracular access to the scheduling functions from \cref{eq:oracles-for-trap-schedule} and an appropriate superposition over $\ket{s}$, $s=0,\ldots,n_t$. To construct superpositions over where $n_t$ is not a power of two, see e.g. \cite[Appendix~J]{su_fault-tolerant_2021}.  
The equivalent construction is necessary for the interaction term together with $f(s)$; see \cref{eq:scheduling-the-traps0}, and encodings for the Coulomb-term in the literature~\cite[Appendix~K]{su_fault-tolerant_2021}. 
There is an argument to make here about the (a)diabaticity of this evolution, which will be part of \cref{app:psuc-discussion}. 

We continue by outlining an LCU encoding of $V^{\rm trap}(s')$ for a specific $0\le s'\le s_1$. 
Coordinate directions are denoted by ${\{x,y,z\}\sim\{0,1,2\}}$ and we use the usual LCU notation of PREP and SEL operations so that ${(\bra{0}_{\rm ancilla}\otimes \mathds{1}_{\rm system}){\rm PREP}^\dagger\cdot{\rm SEL}\cdot{\rm PREP} (\ket{0}_{\rm ancilla}\otimes \mathds{1}_{\rm system}) }$ block-encodes the desired operator.  
The state we are acting on is assumed to be a superposition over $\ket{p}\ket{s'}$, with the state $\ket{p}$ representing a spatial part and $\ket{s'}$ the current step in the evolution of the merging.
The PREP step involves preparing a superposition across nuclei and coordinate directions, 
weighted by the square root of the interaction strength at that point:
\begin{equation}\label{eq:prep-for-trap}
    \ket{0} \ket{p}\ket{s}   \mapsto \sum_{j:{\rm nuclei}} \sum_{s = 0}^{n_t} \sum_{w=0}^2  \frac{\sqrt{\frac{m_j}{2}}\omega_{j,w}}{\sum_{j',w'} \sqrt{\frac{m_{j'}}{2}}\omega_{j',w'}}  \ket{j}  \ket{w} \ket{p}\ket{s} 
\end{equation}
Next, we illustrate the SEL operation, where we omit amplitudes and explicit sums for clarity. The goal here is to compute state-dependent interaction strengths on the fly. 
Namely, we aim to prepare a register that contains $(R_{p_{i,w}} - R_{0;w})^2$, and another containing $g(s)$ to multiply the latter, so that we can use e.g.~\cite{sanders_black-box_2019} to move information from the state into the amplitude.
Consequently,  access to oracles $O_f, O_g$ that represent the scheduling is required, 
\begin{equation}\label{eq:oracles-for-trap-schedule}
    O_{f/g}: \ket{0}\ket{s}\to \ket{f/g(s)} \ket{s}.
\end{equation}
The oracular access to the scheduling function and the label-to-coordinate mapping may be realized with classical data encodings, e.g., a QROM or other techniques~\cite{babbush_encoding_2018,zhang2024circuit,sun2024low}.
Using a number of ancillas given by $n_{\rm anc}$ initialized in zero, and assuming a state that encodes the relative center of the trap $R_0$,
\begin{equation}
    \begin{aligned}
    & \ket{j}\ket{w} \ket{s} \ket{0}_{{\rm a}} \ket{p_1\cdots p_{N_{\rm part}}} \ket{R_{0; w}} \\
    \mapsto & 
    \ket{j} \ket{w} \ket{s} \ket{0}_{{\rm a}'} \ket{g(s)}\ket{R_{p_{j,w}}} \ket{p} \ket{R_{0;w}} \\
    \mapsto &
    \ket{j} \ket{w} \ket{s} \;\;\;\;\;\; \ket{g(s)(R_{p_{j,w}} - R_{0;w})^2} \ket{g(s)}\ket{R_{p_{j,w}}} \ket{p} \ket{R_{0;a, w}} \\
    \mapsto &
    \ket{j} \ket{w} \ket{s} \ket{0}_{{\rm a}''} \ket{g(s)\lvert R_{p_{j}} - R_{0}\rvert^2}  \ket{p} \ket{R_{0;w}}
    \end{aligned}\label{eq:sel-for-trap}
\end{equation}
The size of the necessary ancillary register $\ket{0}_{\rm a}$ that is used to store intermediary and final results scales linearly in the bits of precision used for numerical representation (to store $g(s), R_{p_{j,w}}$ and perform multiplication and sum-of-squares). 
As mentioned above, to finalize the SEL operation, the state information in $\ket{g(s)\lvert R_{p_{j}} - R_{0}\rvert^2}$ needs to be moved into the amplitude. There are multiple ways to do so, such as the inequality-testing approach in \cite{sanders_black-box_2019}, or, the conceptually simplest would be controlled rotations, conditioned on $\ket{g(s)\lvert R_{p_{j}} - R_{0}\rvert^2}$ and applied to the system register. 

We refrain from analyzing exact Toffoli or gate counts as done in \cite{su_fault-tolerant_2021}, expecting that the addition of the trap potentials would not add a significant change here \textit{per single query} to involved oracles and circuits as the appearing arithmetic operations are comparable to the usual potentials in the Hamiltonian. 
The factor that may make the trap potentials more costly and needs to be studied more closely is its sub-normalization factor. This factor captures the number of necessary queries for a successful block-encoding.
We recall the sub-normalization factors from \cite[Appendix K]{su_fault-tolerant_2021},
\begin{equation}\label{eq:subnormalizations-from-yuansu}
    \begin{aligned}
        \alpha_T &\in O\left(\frac{\numel N^{2/3}}{  \lvert\Omega\rvert^{2/3}}\right), & 
        \alpha_V &\in O\left(\frac{\numel^2 N^{1/3}}{\lvert\Omega\rvert^{1/3}}\right), & 
        \alpha_U &\in O\left(\frac{\numel^2 N^{1/3}}{\lvert\Omega\rvert^{1/3}}\right)
    \end{aligned}
\end{equation}
with $T$ kinetic energy, $V$ electron-electron interactions and $U$ nuclear-electron interactions.

The block-encoding factor of the trap potential may be upper-bounded by $\alpha_{\rm trap} \le \lVert \sum_{j\in [\numnuc]} V_{j}^{\rm trap} \rVert$.
Then, $  \alpha_{\rm trap}\le \numnuc m_{\max} \omega_{\max}^2 \max_{r,r'} \lvert r-r'\rvert^2 $. 
In the present discretization, $\max_{r,r'\in\Omega_{\rm trap}} \lvert r-r'\rvert^2 \in O( \lvert \Omega_{\rm trap}\rvert^{2/3})$, where $\Omega_{\rm trap}\subseteq\Omega$ is the part of the domain the trap potential is defined on.
Most of the mass of particles under such a potential will be incentivized to be close to the center; therefore, an upper bound that puts the maximum mass at a maximum distance will not be very tight in most scenarios, and a smaller $\alpha_{\rm trap}$ may be sufficient. 
One could further think about designing the shape of the trap potential to be decaying for larger distances to reduce the encoding cost or oscillatory as in optical tweezers, such that maximum amplitude is always within reach -- we leave concrete specifications up to future work. 
The maximum nuclear mass $m_{\max} = \max_{1\le j \le \numnuc} m_j $ can be treated as a constant factor attached to the nuclei. Then, we have that 
\begin{equation}\label{eq:sub-normalization-trap}
   \alpha_{\rm trap} \in O\left(\omega_{\max}^2 \numnuc\lvert\Omega_{\rm trap}\rvert^{2/3}      \right).
\end{equation}
Up to the frequency $\omega_{\max}$, the sub-normalizations of the present procedure in \cref{eq:subnormalizations-from-yuansu} compared to the trap potential inhibits a few key differences. Because the strength of the trap depends on a maximum distance, there is no notion of `grid density' in the cost. Therefore, the factors in \cref{eq:subnormalizations-from-yuansu} increase when a lower target accuracy is desired. 
The trap encoding is not directly dependent on accuracy, though it is more pronounced on system size. The system size of the trap may be upper-bounded with the overall system studied, as present at the last scattering stage before `evolution' in \cref{fig:framework}.

Therefore, it is key to understand the role of $\omega_{\max}$ to the scattering and whether there is a way to relate this quantity to the system size. 

\subsection{Probability of Success of Mergo-Association}\label{app:psuc-discussion}
To address the choice of trap frequency, we consider that the probability of a diabatic transition into $\ket{a}$ through the mergo-association from \cite{bird2023making,bird2024makingmoleculesmergoassociationrole} can be approximated by the Landau-Zener rule and is proportional to
\begin{equation}
  p_{\rm LZ} \approxeq \exp( -\frac{2\pi}{\hbar}\frac{\omega_{\rm eff}^2}{\lvert \frac{\partial}{\partial s} ( E_{\rm mol} -  E_{\rm atom} ) \rvert }   )
\end{equation}
Here $\omega_{\rm eff}$ is an effective frequency we explain further below that models the strength of the potential. 
Then, we express the rate of change of the energy, $\frac{\partial}{\partial s} ( E_{\rm mol} - \partial E_{\rm atom} )$, by consequence of the chain rule, as $\lvert \partial E_{\rm mol} - \partial E_{\rm atom} \rvert \, v$. Here,   $\partial E$ are energy derivatives near the initial configuration with respect to the internuclear distance and $v$, the speed of the evolution along the scheduling function.
\begin{equation}\label{eq:plz-first-general}
  p_{\rm LZ} \approxeq \exp( -\frac{2\pi}{\hbar}\frac{\omega_{\rm eff}^2}{\lvert \partial E_{\rm mol} - \partial E_{\rm atom} \rvert \, v}   )
\end{equation}
Thus, the success probability of the mergo-association, so that the merging happens without transitioning into a higher vibrational state of the trapped system as desired, is 
\begin{equation}
    p_{\rm suc} \sim 1-p_{\rm LZ}.
\end{equation}
As we pointed out in the previous section, this means that the type of evolution we call successful here is adiabatic. 
It is important to note that this is not adiabatic ground-state preparation, which is explicitly beyond the scope of our work. 
Following another discussion in the preceding section, the threshold to success here considers molecular bound states, so the `upper limit' for adiabaticity would be the highest-lying vibrational bound state. 
One way or another, a practical implementation will need to consider this.
This still poses limits on the speed of evolution (given by $v$ in \cref{eq:plz-first-general}). 
Then,  $v =  \max\{ \lvert\frac{{\rm d}z}{{\rm d} f}\frac{\rm d}{{\rm d}s} f(s^*)\rvert, \lvert\frac{{\rm d}z}{{\rm d} g}\frac{\rm d}{{\rm d}s} g(s^*)\rvert\}$, where $z$ is the considered internuclear distance.

Looking at \cref{eq:plz-first-general}, we can identify that the stronger (``steeper'') the potential as described by the effective frequency, the higher the success, and the faster the merging proceeds, the less likely. This conclusion is consistent with intuition.

We continue by breaking down the quantities that appear.
Following \cite{bird2023making}, $\partial E_{\rm atom}\approx 0$, as it is reasonable to assume the energy surface for separated atoms is flat. 
Taking a harmonic oscillator approximation to the relative motion of the nuclei, \cite[Eq.~(37)]{bird2023making} obtains an approximation to the gradient, 
$\partial E_{\rm mol} \approx \mu\omega^2 z^* $, 
where $z^*$ is the point of contact of the avoided crossing between the two states. In terms of the scheduling functions that we envision, this corresponds to the state at $s=s^*$, $0<s^*<s_0<s_1$ (\cref{fig:trap-int-scheduler}): briefly before both interaction and trap are `fully acting'. 
Assuming an isotropic potential, the spatial coordinate is approximated as 
$z^* =  (\frac{\hbar}{\mu\omega})^{1/2}(3+\frac{\omega_a}{\omega})^{1/2}$ so that~\cite[Eq.(35)]{bird2023making}
\begin{equation}\label{eq:expression-energy-derivative}
    \partial E_{\rm mol}\approx({\hbar\mu})^{1/2}\omega(3\omega+\omega_a)^{1/2} ,
\end{equation}
where $\omega_a$ is defined as the frequency associated with the bonded vibrational excited state.

To estimate $\omega_{\rm eff}$, we can also follow \cite[Eq.~(52)]{bird2023making}:
\begin{align}
    \omega_{\rm eff}^2 
    &\approxeq 
    \left(\frac{\langle a \vert (\mathds{1} - \op{000}) V^{\rm int} \vert 000 \rangle}{1 - \lvert \langle a \vert 000 \rangle \rvert^2}\right)^2 \nonumber\\
    &\approx \langle a \vert V^{\rm int} \vert 000 \rangle^2
    =
    \frac{2\hbar^2}{\sqrt\pi}\omega_a^{1/2}\omega^{3/2} \exp[-\frac{1}{2}\left(3+\frac{\omega_a}{\omega}\right)] \label{eq:approx-omega-eff}.
\end{align}
States ``000'' describe the vibrational ground-state and ``$a$'' an excited vibrational state, $\omega$ is the frequency of the harmonic oscillator trap.
The approximation to only look at the transition element itself is investigated in \cite[Fig.~6]{bird2023making} -- for a merely qualitative argument, this is a sufficient choice.
The choice for $\omega_a$ in our case is an estimate for an upper threshold of the respective \textit{bonded} vibrational energy subspace.
Introducing a `binding energy'-quantity $E_a = \hbar \omega_a$, we can rewrite $\omega_{\rm eff}^2$ to be approximately proportional to \cite[Eq.~(54)]{bird2023making}, 
\begin{equation}\label{eq:omega-thru-bonding}
    \omega^2_{\rm eff}\appropto 
    \omega^{3/2}({E_a}/{\hbar})^{1/2} \exp(-\frac{E_a}{\hbar\omega}) 
    =
    \omega^2 \tilde E_a^{1/2}\exp(-\tilde E_a).
\end{equation}
In the latter equation, we introduce the relative binding energy $\tilde E_a = E_a/(\hbar\omega)$, defined in relation to the trap energy.
Then, 
\begin{align}\label{eq:psuc-for-mergo-association-after}
    p_{\rm LZ} &\approx {\rm exp}\left(
        -4\left({\frac{\hbar\pi}{\mu}}\right)^{1/2}
        \left(\frac{\omega\omega_a}{3\omega+\omega_a}\right)^{1/2}
        \frac{e^{-\frac{1}{2}\left(3+\frac{\omega_a}{\omega}\right) }  }{v} 
    \right) 
    \nonumber
    \\
    &=
    \exp(
    -4\left(\frac{\pi}{\mu}\right)^{1/2}
    \left(
    \frac{
        \hbar \omega E_a
    }{
    3\hbar\omega + E_a
    }
    \right)^{1/2}
    \frac{
    \exp(-\frac{1}{2}\frac{E_a}{\hbar\omega}-\frac{3}{2})
    }{v}
    )
    \nonumber
    \\
    &=
    \exp(
    -4\left(\frac{\pi}{\mu}\right)^{1/2}
    \left( 
    \hbar\omega
    \frac{
        \tilde E_a
    }{
    3 + \tilde E_a
    }
    \right)^{1/2}
    \frac{
    \exp(-\frac{1}{2}\tilde E_a - \frac{3}{2})
    }{v}
    )
    .
\end{align}

\section{Construction of Measurement Oracles for Certifying Reactions}\label{app:meas-oracles-appendix}
For this oracle construction, we assume a first-quantized real space implementation. 
The choice of representation in first-quantization on real space makes the construction of the following oracle easier; of course,  plane-wave or plane-wave dual bases~\cite{babbush_low-depth_2018} are possible too and can be translated to with appropriate transformations.
The encoded quantum states are linear combinations of wavefunctions with Fermionic (antisymmetric) and Bosonic (symmetric) parts.
Within such a combination, a single component looks like a tensor product over electronic $\{r\}\sim\{r_j\}_{j=1}^{\numel}$ and nuclear $\{R\}\sim \{R_k\}_{k=1}^{\numnuc}$ grid labels (represented on the grid \cref{eq:grid-labelling}),
\begin{align}
    \left| \psi (\{r, R\}) \right> = 
    \ket{\mathtt{r}_1\cdots \mathtt{r}_{\numel}}_{\mathrm{el}}\ket{\mathtt{r}_{1}\cdots \mathtt{r}_{\numnuc}}_{\mathrm{nuc}} ; 
    \label{eq:position_basis_state}
\end{align}
We further remember that the above collapsed notation that also contain spin information in an extra qubit, i.e., $\ket{\mathtt r_j}\sim \ket{\mathtt r_j}\otimes\ket{\sigma_j}$ with $\sigma_j\in\{\uparrow, \downarrow\}$, and analogous for nuclei if desired.
Then, a density matrix formulation of linear combinations of such a state is
\begin{align}
    \rho = \sum_{\{r,R\}, \{r',R'\}} \rho_{\{r,R\}, \{r',R'\}} \op{ \psi (\{r,R\}) }{\psi (\{r',R'\})},
    \label{eq:position_basis_mixed_state}
\end{align}
with the entries of the density matrix in the position basis, $\rho_{\{r,R\}, \{r',R'\}} \in \mathbb{C}$. 
The goal of this section is to construct an oracle that can distinguish the `good', reacted components of such a state from the `bad', non-reacted components. 
Given a state representing two molecular fragments that have evolved for some time, we want to distinguish the parts of the wavefunction that correspond to the fragments having reacted to form a single molecule from those where they are still separate unbound fragments.

As we consider linear transformations, it suffices to study the effect on the individual components from \cref{eq:position_basis_state} to draw conclusions for a general state in \cref{eq:position_basis_mixed_state}. 
Therefore, we first focus on a single position-basis state $\left| \psi (\{r,R\}) \right>$ in the upcoming \cref{subsec:geom-crit-oracles} before we consider exchange symmetry in \cref{subsec:distinguishable-oracles} and more general superpositions in \cref{subsec:oracle_on_superpositions}.

\subsection{Geometrical Criteria}\label{subsec:geom-crit-oracles}
We now discuss an approach to test whether or not a molecular bond has formed for a state of the form $\left| \psi (\{r,R\}) \right>$. When atomic nuclei can be precisely located, a reasonable description for the molecular structure can be given in terms of a set of inter-nuclear distances $\{R^{\text{react}}_{jk}\}$. Using this information, a corresponding criterion $\mathcal C$ for whether a given configuration corresponds to a desired molecule can then be constructed,
\begin{align}
    \mathcal{C}_{\text{geom}, \varepsilon} (\{r,R\}) ) = \begin{cases} 1, & \left| \left\|R_j - R_k\right\| - R^{\text{react}}_{jk} \right| \leq \varepsilon \; \forall j,k\\
    0, & \text{else.}
    \end{cases}
\end{align}
Note that evaluating this criterion is classically efficient with respect to the number of bits devoted to the constituent quantities, scaling no worse than $O(N_{\text{nuc}}^2)$ in the number of nuclei and no worse than quadratically in the bit-precision due to a product and a square root in the evaluation of the norm. In other words, evaluating this function is classically efficient in the size of the molecule as long as the bit-precision only grows polynomially, corresponding to the requirement that the number of grid points in the simulation should not grow faster than exponential in the involved particle number. 

In some cases, it may be possible to devise more accurate criteria, such as by comparing the angles between inter-nuclear distances to desired bond angles. On the other hand, there may be cases where less is known about the target structure. Then, the best that one can hope for is to check that the two fragments are at least spatially adjacent, corresponding to the looser requirement
\begin{align}
    \tilde{\mathcal{C}}_{\text{geom}, \Delta} (\{r,R\}, I)  = \begin{cases} 1, & \left\|R_j - R_k\right\| \leq \Delta_{jk} \; \forall (j,k)\in I\subseteq[\numnuc]\times[\numnuc]\\
    0, & \text{else.}
    \end{cases}
\end{align}
for some suitable $\Delta_{jk} >0$ and a set of locations of interest $\mathcal I$, in which we only check a subset of the encoded locations regarding proximity. 
There are many ways this type of geometric requirements could be mixed and matched, e.g., utilizing different levels of knowledge or different precisions $\varepsilon$ for different groups of atoms. Either way, common to these approaches is the fact that the classical evaluation of the criterion is efficient under the mild restrictions to the grid resolution outlined above.

A result of this classical efficiency is that a corresponding unitary to perform the computation can also be implemented efficiently on a quantum computer using reversible logic. Specifically, for a given efficient criteria $\mathcal{C}$ of the form considered here, the following unitary can be implemented:
\begin{equation}
U_{\mathcal{C}} \ket{p_{\numel}\cdots p_{\numel+\numnuc}}_{\mathrm{nuc}}\ket{0}
=  \ket{p_{\numel}\cdots p_{\numel+\numnuc}}_{\mathrm{nuc}}\ket{\mathcal{C}(\{r,R\})}.
\end{equation}   
Thus, extracting information about the criteria into an ancilla qubit for easy access is possible. Using a measurement of this ancilla then allows for a projection onto the good ($\mathcal{C}=1$) subspace, heralded by an outcome ``1'' of the measurement. However, the outcome ``0'' similarly fully projects the state into the bad subspace ($\mathcal{C}=0$). Furthermore, care needs to be taken to avoid either kind of projection destroying physically important symmetries of the state, including Fermionic and Bosonic particle-exchange symmetries. Mitigating each of these potential problems will be the topics of the following two sections. 

\subsection{Preserving Exchange Symmetry Throughout Measurement}\label{subsec:distinguishable-oracles}
So far, we have been primarily considering a single component of the wavefunction, $ \left| \psi (\{r,R\}) \right>$, in which the positions of all particles are fully specified. However, we typically deal with a collection of nuclei and electrons modeled through wavefunctions that need to be of the correct symmetry with respect to the exchange of particles. 
We focus on preserving exchange symmetry compared to other symmetries present in the system that may also be broken in the real physical processes. 
Hence, the goal is to evaluate reaction criteria in a manner that does not violate the symmetries of present states.

\subsubsection{Symmetrization of the wavefunction}
Recall that the wavefunctions we encode are given as superpositions of position-labelling basis states, 
\begin{equation}\label{eq:position-basis-without-symmetry}
    \left(|{\mathtt{r}_1}\rangle |{\mathtt{r}_2}\rangle \cdots |{\mathtt{r}_{\numel}}\rangle\right)\left(|{\mathtt{R}_{1}}\rangle|{\mathtt{R}_{2}}\rangle \cdots |{\mathtt{R}_{\numnuc}}\rangle\right)
\end{equation}
We repeat that by $\ket{\mathtt{r}_j}, \ket{\mathtt{R}_k}$ we mean the labels for grid points and spin rather than the resolved coordinates and choose this notation as it is more illustrative in the context of (anti)symmetrization of the wavefunctions.
For Fermionic (indistinguishable) particles, the wavefunction needs to be anti-symmetrized with respect to the indistinguishable degrees of freedom. Formally, this can be done by applying the antisymmetrization operator $\mathcal{A}$. Using the electronic degrees of freedom as an example, and with $\sigma$ a permutation from the permutation group over $\numel$ elements, $S_{\numel}$:
\begin{align}\label{eq:2-Fermions-and-another}
    \mathcal{A}\left(\ket{\mathtt{r}_1}\ket{\mathtt{r}_2} \cdots \ket{\mathtt{r}_{\numel}}\right) 
    &= \frac{1}{\sqrt{\numel!}} \sum_{\sigma\in S_{\numel}} \text{sgn}(\sigma) \ket{\mathtt{r}_{\sigma(1)}} \ket{\mathtt{r}_{\sigma(2)}}\cdots\ket{\mathtt{r}_{\sigma(\numel)}}
\end{align}
Similarly, Bosonic degrees of freedom will be symmetric with respect to exchange, described by symmetrization $\mathcal S$ of the general form 
\begin{align}
    \mathcal S \left(\ket{\mathtt{R}_{1}}\ket{\mathtt{R}_{1}}\cdots \ket{\mathtt{R}_{\numnuc}}\right) 
    &=
    \frac{1}{\sqrt{k!}} \sum_{\sigma\in S_{\numnuc}}  \ket{\mathtt{R}_{\sigma(1)}} \ket{\mathtt{R}_{\sigma(2)}}\cdots\ket{\mathtt{R}_{\sigma(\numnuc)}}
\end{align}
In general, a set of nuclei to be modeled as indistinguishable will correspond to a subset of nuclear registers to be (anti)symmetrized. Thus, the desired symmetry properties are characterized by sets of registers to be symmetrized, $\left\{ B_i\right\}_{i=1}^{N_B}$, and sets of registers to be anti-symmetrized, $\left\{F_i\right\}_{i=1}^{N_F}$. Denoting which basis states an operator acts on by a subscript, one gets the following anti-symmetrized state:
\begin{align}
    \mathcal{A}\left(\ket{\mathtt{r}_1} \ket{\mathtt{r}_2} \cdots \ket{\mathtt{r}_{\numel}} \right)  \otimes \left( \prod_{i=1}^{N_F} \mathcal A_{F_i} \prod_{j=1}^{N_B} \mathcal S_{B_j}  \right) \left(\ket{\mathtt{R}_1}\ket{\mathtt{R}_2} \cdots \ket{\mathtt{R}_{\numnuc}} \right). \label{eq:symmetrized_state}
\end{align}

\subsubsection{Illustrative example: H${}_2$O${}_2$}
To elucidate our notation, consider the molecule H${}_2$O${}_2$, and assume that the non-symmetrized positions are encoded as
\begin{align}
    \ket{\mathtt R_{O,1}}\ket{\mathtt R_{O,2}}\ket{\mathtt R_{H,1}}\ket{\mathtt R_{H,2}} . \label{eq:non-symmetrized_H2O2}
\end{align}
In this case, the oxygen nuclei are spin-0 nuclei, meaning they should be symmetrized. On the other hand, the hydrogen nuclei are spin-1/2 particles, meaning they should be anti-symmetrized. We can represent this by the following two sets:
\begin{align}
    B_1 = \{ 1,2\}, \quad &
    F_1 = \{ 3,4\}.
\end{align}
The first tells us to symmetrize the oxygen registers (1 and 2), and the second tells us to anti-symmetrize the hydrogen registers (3 and 4). The state after symmetrization is therefore
\begin{align}
    \mathcal A_{F_1} \mathcal S_{B_1} (\ket{\mathtt R_{O,1}}\ket{\mathtt R_{O,2}}\ket{\mathtt R_{H,1}}\ket{\mathtt R_{H,2}}) 
    =& 
    \frac{1}{2} \left( \ket{\mathtt R_{O,1}}\ket{\mathtt R_{O,2}} + \ket{\mathtt R_{O,2}}\ket{\mathtt R_{O,1}}\right) \nonumber\\
    &\otimes \left( \ket{\mathtt R_{H,1}}\ket{\mathtt R_{H,2}} - \ket{\mathtt R_{H,2}}\ket{\mathtt R_{H,1}} \right).
\end{align}

\begin{figure}[t]
    \centering
    \includegraphics[width=0.55\linewidth]{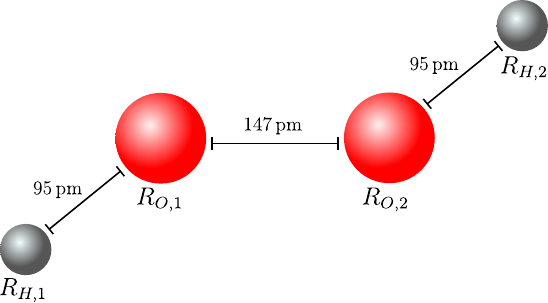}
    \caption{Equilibrium configuration of a H${}_2$O${}_2$ molecule, with oxygen nuclei marked in red and hydrogen nuclei marked in grey.}
    \label{fig:H2O2}
\end{figure}

The encoded molecular state will generally be a superposition of states of form in~Eq. \eqref{eq:symmetrized_state}, i.e., the tensor products of an antisymmetric wavefunction, a symmetric wavefunction, and possibly a non-symmetrized component. Then, for the remainder, it suffices to discuss the oracle's effect on one term of the superposition. To see what symmetry implies for the construction of oracles, consider again the example of H${}_2$O${}_2$. Looking at the form of the wavefunction in \cref{eq:non-symmetrized_H2O2} and the equilibrium geometry of the molecule, one might be tempted to define a criteria of the form
\begin{align}
    \mathcal{C}_\varepsilon(\{R\}) =  \Big(|\norm{R_1 - R_3} - \SI{95}{\pico \metre}| < \varepsilon) &\land (|\norm{R_2 - R_4} - \SI{95}{\pico \metre}| < \varepsilon) \nonumber\\
    &\land \, (|\norm{R_1 - R_2} - \SI{147}{\pico \metre}| < \varepsilon)\Big)
\end{align}
in suitable numerical units and with subscripts denoting the ordering of the registers. After all, if we achieve the equilibrium configuration (See \cref{fig:H2O2}), an oracle based on this criteria will, by construction, certify the non-symmetrized state as reacted:
\begin{align}
    U_{\mathcal{C}} \ket{\mathtt R_{O,1}}\ket{\mathtt R_{O,2}}\ket{\mathtt R_{H,1}}\ket{\mathtt R_{H,2}} \ket{0} =& \ket{\mathtt R_{O,1}}\ket{\mathtt R_{O,2}}\ket{\mathtt R_{H,1}}\ket{\mathtt R_{H,2}} \nonumber\\
    & \; \otimes \ket{\mathcal{C}_\varepsilon(\{R_{O,1}, R_{O,2}, R_{H,1}, R_{H,2}\})} \nonumber\\
    =& \ket{\mathtt R_{O,1}}\ket{\mathtt R_{O,2}}\ket{\mathtt R_{H,1}}\ket{\mathtt R_{H,2}} \ket{1}
\end{align}
However, the same cannot be said for states where the positions are permuted. For instance, swapping the oxygen positions, the evaluation of the oracle reads
\begin{align}
    \mathcal{C}_\varepsilon(\{R_{O,2}, R_{O,1}, R_{H,1}, R_{H,2}\}) = & \, (|\norm{R_{O,2} - R_{H,1}} - \SI{95}{\pico \metre}| < \varepsilon) \nonumber\\
    &\land (|\norm{R_{O,1} - R_{H,2}} - \SI{95}{\pico \metre}| < \varepsilon) \nonumber\\ & \land \, (|\norm{R_{O,2} - R_{O,1}} - \SI{147}{\pico \metre}| < \varepsilon)
\end{align}
and consulting the geometric configuration in \cref{fig:H2O2}) shows that this Boolean formula evaluates to zero. Thus,
\begin{align}
    U_{\mathcal{C}} \ket{\mathtt R_{O,2}}\ket{\mathtt R_{O,1}}\ket{\mathtt R_{H,1}}\ket{\mathtt R_{H,2}} \ket{0} = \ket{\mathtt R_{O,2}}\ket{\mathtt R_{O,1}}\ket{\mathtt R_{H,1}}\ket{\mathtt R_{H,2}} \ket{0}.
\end{align}
Considering the full symmetrized wavefunction, one finds
\begin{align}
    U_{\mathcal{C}}  \left[ \frac{1}{2}\left( \ket{\mathtt R_{O,1}}\ket{\mathtt R_{O,2}} + \ket{\mathtt R_{O,2}}\ket{\mathtt R_{O,1}}\right) \otimes \left( \ket{\mathtt R_{H,1}}\ket{\mathtt R_{H,2}} - \ket{\mathtt R_{H,2}}\ket{\mathtt R_{H,1}} \right) \right] \ket{0} \nonumber \\
    = \frac{1}{2} \left[ \ket{\mathtt R_{O,1}}\ket{\mathtt R_{O,2}} \ket{\mathtt R_{H,1}}\ket{\mathtt R_{H,2}} - \ket{\mathtt R_{O,2}}\ket{\mathtt R_{O,1}} \ket{\mathtt R_{H,2}}\ket{\mathtt R_{H,1}} \right] \ket{1} \nonumber\\
    + \frac{1}{2} \left[ \ket{\mathtt R_{O,2}}\ket{\mathtt R_{O,1}} \ket{\mathtt R_{H,1}}\ket{\mathtt R_{H,2}} - \ket{\mathtt R_{O,1}}\ket{\mathtt R_{O,2}} \ket{\mathtt R_{H,2}}\ket{\mathtt R_{H,1}} \right] \ket{0}
\end{align}
Despite all four components corresponding to the same molecule (just with particle labeling altered), only two of them have been certified by the oracle. Furthermore, if the ancilla is measured, the resulting projected state is no longer of the symmetric form described in \cref{eq:symmetrized_state}, and direct inspection shows that it is neither symmetric nor anti-symmetric upon permutations among the nuclear coordinates. 

One possible response to this problem might be simply re-applying an (anti-)symmetrization step whenever a projection occurs. In principle, this operation scales linearly (up to logarithmic factors) in the system size\cite{berry_improved_2018, liu2024low}; thus, it would not change the polynomial runtime of the algorithm. However, existing algorithms require a particular structure on the input state, a requirement that is linked to unitarity~\cite{berry_improved_2018} and thus likely challenging to lift. Furthermore, even with a hypothetical generalization of the algorithm, the present work proposes a quasi-continuous weak-measurement strategy for monitoring the reaction criteria. Since this entails frequent (weak) measurements during the simulation run, it is likely to lead to a high total number of measurements, scaling extensively with the simulation time. Restoring symmetry after each measurement could, therefore, add significant complexity to the algorithm, even if a fast and applicable symmetrization algorithm is available.
Therefore, we next discuss a more comprehensive approach to address these shortcomings.

\subsubsection{Symmetry of states}
To build certification oracles that are valid under the correct exchange symmetry, we will first formalize the symmetrization requirement for the states of the system. Consider first a single set of identical particles, represented by a set of indices $\mathcal X$. The group of permutations of this set is characterized by the symmetric group, $S_{|\mathcal X|}$, and for each permutation $\sigma \in S_{|\mathcal X|}$, we can define a unitary consisting of SWAP gates that implements the permutation our registers:
\begin{align}
    U_{\sigma} \ket{\mathtt R_1}\ket{\mathtt R_2} \dots \ket{\mathtt R_{|\mathcal X|}} = \ket{\mathtt R_{\sigma(1)}}\ket{\mathtt R_{\sigma(2)}} \dots \ket{\mathtt R_{\sigma(|\mathcal X|)}} ,
\end{align}
or, more compactly
\begin{align}
    U_{\sigma} \ket{\psi \{ r, R\} } = \ket{\psi \{\sigma [ r, R\}] }
\end{align}
$\sigma [\{r,R\}]$ denotes the (ordered) set of positions after the permutation has been applied. In a group-theory language, this corresponds to a representation of the group $S_{|\mathcal X|}$ onto the space of states. In terms of these constructions, the symmetrization requirement is
\begin{align}
    U_{\sigma} \ket{\psi} &= \begin{cases} \text{sgn}(\sigma) \ket{\psi} &  \text{Fermions}\\
    \ket{\psi} &  \text{Bosons}
    \end{cases} &  \forall \sigma \in S_{|\mathcal X|}
\end{align}
where $\text{sgn}(\sigma)$ is the sign of the permutation; the expression depends on whether the identical particles represented by the indices in $\mathcal X$ are Fermions or Bosons.

Generalizing this construction to multiple sets of Bosonic and Fermionic particles, the symmetry group takes the form of a product group of the individual Bosonic and Fermionic symmetries,
\begin{align}
    S &= S_B \otimes S_F \\
    S_B &= \bigotimes_{j=1}^{N_B} S_{|B_j|}, \quad 
    S_F = \bigotimes_{i=1}^{N_F} S_{|F_i|}  ,
\end{align}
with the group representation taking a similar product form. Note that the tensor product structure implies that every element in $S$ can be decomposed into the product of a Fermionic and Bosonic component, $\sigma = \sigma_B \otimes \sigma_F$. Using this notation, the symmetry requirement becomes
\begin{align}
    U_{\sigma} \ket{\psi} = \text{sgn} (\sigma_F) \ket{\psi} \quad \forall \sigma  \in S.
\end{align}

\subsubsection{Symmetry of oracles}
Using the notation introduced above, we can now state the requirement that a reaction criteria $\mathcal{C}$ needs to fulfill to respect the symmetry defined in the previous section. Specifically, we will say that a criterion $\mathcal{C}$ respects symmetry when a projective measurement of $\mathcal{C}$ on a symmetrized state produces post-measurement states that are still correctly symmetrized. Below, we will show that this property holds if and only if
\begin{align}
    \mathcal{C}( \sigma [\{r,R\}]) = \mathcal{C}( \{r,R\}) \quad \forall \sigma \in S, \; \forall \{r,R\} \; .
    \label{eq:symmetric_criteria}
\end{align} 
To see that this is sufficient criteria, consider the projection operators related to the readout of $\mathcal{C}$ using an ancilla qubit:
\begin{align}
    \Pi_{\pm}^{\mathcal{C}} = \frac{1}{2}  U_{\mathcal{C}}^\dagger  \; (\mathds{1} \otimes \left( \mathds{1} \pm Z \right)_{\text{anc}}) \;  U_{\mathcal{C}} \\
    \tilde{\Pi}_{\pm}^{\mathcal{C}} = ( \mathds{1} \otimes \bra{0}_{\text{anc}} ) \; \Pi_{\pm}^{\mathcal{C}} \; ( \mathds{1} \otimes \ket{0}_{\text{anc}})
\end{align}
Looking at the constituent operators, we can note that
\begin{align}
    U_{\mathcal{C}} \, (U_{\sigma} \otimes \mathds{1}) \ket{\psi (\{r,R\})} \ket{0} &= U_{\mathcal{C}}  \ket{\psi ( \sigma[\{r,R\}])} \ket{0} \nonumber\\
    &=  \ket{\psi ( \sigma[\{r,R\}])} \ket{\mathcal{C}( \sigma[\{r,R\}])}\\
    (U_{\sigma} \otimes \mathds{1}) \, U_{\mathcal{C}} \ket{\psi (\{r,R\})} \ket{0} &= U_{\sigma}  \ket{\psi (\{r,R\})} \ket{\mathcal{C}( \{r,R\})} \nonumber\\
    &=  \ket{\psi ( \sigma[\{r,R\}])} \ket{\mathcal{C}( \{r,R\})}  .
\end{align}
Thus, if the relation in Eq. \eqref{eq:symmetric_criteria} holds, it implies that
\begin{align}
    \left[ U_{\mathcal{C}} ,(U_{\sigma} \otimes \mathds{1}) \right] = \left[ U_{\mathcal{C}}^\dagger , (U_{\sigma} \otimes \mathds{1}) \right] = 0 \quad \forall \sigma \in S .
\end{align}
Therefore, the projectors also commute with the permutations,
\begin{align}
   \left[ \Pi_{\pm}^{\mathcal{C}}, (U_{\sigma} \otimes \mathds{1}) \right] = \left[ \tilde{\Pi}_{\pm}^{\mathcal{C}}, U_{\sigma}  \right] = 0 \quad \forall \sigma \in S ,
\end{align}
which in turn means the post-projection states of (anti)symmetrized states are also (anti)symmetrized
\begin{align}
     U_{\sigma}  \left( \tilde{\Pi}_{\pm}^{\mathcal{C}} \ket{\psi} \right) &=  \tilde{\Pi}_{\pm}^{\mathcal{C}} \, U_{\sigma}  \ket{\psi}\nonumber\\
     &=  \tilde{\Pi}_{\pm}^{\mathcal{C}}\,  \text{sgn} (\sigma_F) \ket{\psi} \nonumber\\
     &= \text{sgn} (\sigma_F) \left( \tilde{\Pi}_{\pm}^{\mathcal{C}} \ket{\psi}\right) .
\end{align}
Hence, adhering to Eq. \eqref{eq:symmetric_criteria} is sufficient for the measurements of the criteria to preserve (anti)symmetry. To see that it is also a necessary condition, assume that a configuration $\{r_0, R_0\}$ and permutation $\sigma_0$ exist so that
\begin{align}
    \mathcal{C}( \sigma [\{r_0,R_0\}]) \neq \mathcal{C}(  \{r_0, R_0\}).
    \label{eq:symmetry-breaking_criteria}
\end{align}
However, suppose for contradiction that the projectors related to $\mathcal{C}$ still map (anti)symmetrized states to (anti)symmetrized states. 
Under this assumption, consider a state $\ket{\psi}$ formed by the (anti)symmetrization of $\ket{\psi(\{r_0,R_0\})}$. By definition, this state has a non-zero overlap with $\ket{\psi(\{r_0,R_0\})}$ 
Writing the decomposition of the state in terms of positions,
\begin{align}
    \ket{\psi} = \sum_{r,R} \alpha_{\{r,R\}} \ket{\psi(\{r,R\})}  ,
\end{align}
this is equivalent to the statement that $\abs{\alpha_{\{r_0,R_0\}}}^2 > 0$.

Assume now without loss of generality that $\mathcal{C}(  \{r_0, R_0\}) = 0$, and consider 
\begin{align}
    U_{\sigma_0} \tilde{\Pi}_{+}^{\mathcal{C}} \ket{\psi} = \text{sgn} (\sigma_{0,F})  \tilde{\Pi}_{+}^{\mathcal{C}} \ket{\psi}
\end{align}
which is fulfilled by the assumption that the projectors preserve (anti)symmetry. Since $\left(\tilde{\Pi}_{+}^{\mathcal{C}}\right)^2 = \tilde{\Pi}_{+}^{\mathcal{C}}$\footnote{To see this in the case of the tilded operators, compare their effects on states from the complete basis $\{\ket{\psi (\{ r, R \})} \}$.}, this implies
\begin{align}
       \tilde{\Pi}_{+}^{\mathcal{C}} U_{\sigma_0} \tilde{\Pi}_{+}^{\mathcal{C}} \ket{\psi} = \text{sgn} (\sigma_{0,F})  \tilde{\Pi}_{+}^{\mathcal{C}} \ket{\psi} .
\end{align}
Therefore, the norms of the left-hand and right-hand sides above are also equal.  Using the following sets
\begin{align}
    A &= \{ \{r, R\} : C(\{r, R\}) = 0 \}\\
    B_{\sigma_0} &= \{ \{r, R\} : C(\sigma_0[\{r, R\}]) = 0 \}
\end{align}
this equality of norms can be written as
\begin{align}
    \sum_{ \{r, R\} \, \in \, A \cap B_{\sigma_0}} \abs{\alpha_{\{r,R\}}}^2 = \sum_{ \{r, R\} \, \in \, A} \abs{\alpha_{\{r,R\}}}^2 
\end{align}
Note that basic set theory implies that $A \cap B_{\sigma_0} \subseteq A$, meaning every term in the left-hand sum is also present in the right-hand sum. Subtracting these common terms, this implies
\begin{align}
 \sum_{ \{r, R\} \, \in \, A \backslash B_{\sigma_0}} \abs{\alpha_{\{r,R\}}}^2 = 0
\end{align}
There can be no nonzero amplitude for any configurations in $A \backslash B_{\sigma_0}$. However, by assumption $\mathcal{C}(  \{r_0, R_0\}) = 0$, meaning $\{r_0, R_0\} \in A$. Furthermore, Eq. \eqref{eq:symmetry-breaking_criteria} implies that $C(\sigma_0[\{r, R\}]) \neq 0$, meaning $\{r_0, R_0\} \notin B_{\sigma_0}$. Thus, $\{r_0, R_0\} \in A \backslash B_{\sigma_0}$ and $\abs{\alpha_{\{r_0,R_0\}}}^2 > 0$. We have, in other words, arrived at a contradiction. For a criteria $\mathcal{C}$ that does not fulfill \cref{eq:symmetric_criteria}, configurations necessarily exist where measuring $\mathcal{C}$ breaks symmetrization.

Note that all arguments above apply equally to other groups of spatial transformations under which the wave function should be invariant (up to a phase), for instance, SO(3). They also may be extended to other operators that commute with the Hamiltonian, such as total spin. In general, one should construct criteria that are maximally invariant under trivial transformations. One will otherwise discard components that correspond to desired configurations, yielding artificially low success probabilities (e.g., 50\% in the case of fully formed H${}_2$O${}_2$).

\subsection{Effect of Oracle Measurement for General Superpositions}\label{subsec:oracle_on_superpositions}
In this section, we cover the effect of weak measurement of an oracle on a general mixed state. The goal is to elucidate the measurement procedure, the effects of the measurements, and the role of the parameter $\delta$ describing the strength of the measurement.

We recall the form of a general input state,
\begin{align}
    \rho = \sum_{j,k} \rho_{j,k} \op{\psi(\{r,R\}_j)}{\psi(\{r,R\}_k)}
\end{align}
as well as access to a classical criterion $\mathcal{C}$ and a unitary implementation $U_{\mathcal{C}}$, as discussed in Sec. \ref{subsec:geom-crit-oracles}. 
Further, we recall the definition of the set of `successful states' marked by $A$ and `unsuccessful states' marked by $B$ from the main text as:
\begin{align}
    A &= \{ j : \mathcal{C}(\{r,R\}_j)=1 \} \nonumber\\
    B &= \{ j : \mathcal{C}(\{r,R\}_j)=0 \} \nonumber\\
    \ket{\psi_j} :&=  \ket{\psi(\{r,R\}_j)}
\end{align}
Then, the input state from above can be rewritten to account for the `successful' and `unsuccessful' subspaces as
\begin{align}
    \rho =& \sum_{j,k \in A} \rho_{j,k} \op{\psi_j}{\psi_k} + \sum_{j,k \in B} \rho_{j,k} \op{\psi_j}{\psi_k} \nonumber\\
    &+ \sum_{j \in A, k \in B} \left( \rho_{j,k} \op{\psi_j}{\psi_k} + \rho_{k,j} \op{\psi_k}{\psi_j} \right). \label{eq:input_to_oracle2}
\end{align}
Below, we will go through each step outlined in the main text in \cref{enumerate-here}, discussing motivation and relevant considerations.

\subsubsection{Step 1: First application of $U_{\mathcal{C}}$}
This step aims to extract the information represented by the oracle into an ancilla for easy access. Using the definition of the sets $A$ and $B$, the state resulting from this step can be explicitly written as
\begin{align}
    U_{\mathcal{C}} \left( \rho \otimes \op{0} \right) U_{\mathcal{C}}^\dagger =& \sum_{j,k \in A} \rho_{j,k} \op{\psi_j}{\psi_k} \otimes \op{1} + \sum_{j,k \in B} \rho_{j,k} \op{\psi_j}{\psi_k} \otimes \op{0} \nonumber\\
    &+ \sum_{j \in A, k \in B} \left( \rho_{j,k} \op{\psi_j}{\psi_k} \otimes \op{1}{0} + \rho_{k,j} \op{\psi_k}{\psi_j} \otimes \op{0}{1} \right).
\end{align}
At this stage, a measurement of the ancilla qubit would project the state onto either the first or the second term in this sum. 
In the case of projection onto the first term (i.e., a measurement outcome of '1'), this projection would correspond to a heralded projection onto the desired set of states, as identified by the oracle. However, such a measurement would come with a risk of a projection onto the space that does not correspond to success. Thus, a full measurement risks projecting the state out of the desired space, and repeated measurements would risk a Zeno-effect-like freezing of the dynamics in the $B$ subspace.
\subsubsection{Step 2+3: Preparation for weak measurement}
To avoid the problems discussed above, this step \emph{partially} extracts the information stored in the first ancilla into a second ancilla, then resets the first ancilla using a second application of $U_{\mathcal{C}}$. For ease of notation, we combine these two steps here and omit the first ancilla (now unentangled and in the state $\op{0}{0}$) from the expression:
\begin{align}
    \rho_{\delta} =& \left( \cos(\delta)^2 \sum_{j,k \in A} \rho_{j,k} \op{\psi_j}{\psi_k} + \sum_{j,k \in B} \rho_{j,k} \op{\psi_j}{\psi_k} + \cos(\delta) \sum_{j \in A, k \in B} \left( \rho_{j,k} \op{\psi_j}{\psi_k} + \rho_{k,j} \op{\psi_k}{\psi_j} \right) \right) \otimes \op{0}{0} \nonumber\\
    &+ \sin(\delta) \cos(\delta) \sum_{j,k \in A} \rho_{j,k} \op{\psi_j}{\psi_k} \otimes \left( \op{0}{1} + \op{1}{0}\right) \nonumber\\
    &+ \sin (\delta) \sum_{j \in A, k \in B} \left( \rho_{j,k} \op{\psi_j}{\psi_k} \otimes \op{1}{0} + \rho_{k,j} \op{\psi_k}{\psi_j} \otimes \op{0}{1} \right) \nonumber\\
    &+ \sin(\delta)^2 \sum_{j,k \in A} \rho_{j,k} \op{\psi_j}{\psi_k} \otimes \op{1}{1}.
\end{align}
Note that only the first and last of these expressions will be relevant once the ancilla is measured.
\subsubsection{Step 4: Measurement and heralding}
We are now ready to see the effect of a measurement of the second ancilla and to understand the role played by the parameter $\delta$. Consider first the projectors corresponding to the two measurement outcomes of the projective measurement:
\begin{align}
    \Pi_A = \mathds{1} \otimes \left( \frac{\mathds{1} - Z}{2} \right)\\
    \Pi_B = \mathds{1} \otimes \left( \frac{\mathds{1} + Z}{2} \right)   
\end{align}
Assume now that the measurement yields a '1' outcome. The probability of this event is 
\begin{align}
    p_1 &= \text{Tr}( \rho_{\delta} \Pi_A ) = \text{Tr}( \Pi_A \rho_{\delta} \Pi_A ) \nonumber\\
    &= \sin(\delta)^2 \, \text{Tr}\left( \sum_{j,k \in A} \rho_{j,k} \op{\psi_j}{\psi_k} \right) \nonumber \\
    &= \sin(\delta)^2 \sum_{j \in A} \rho_{j,j} \nonumber\\
    &= \sin(\delta)^2 \, p_{\text{suc}} \; ,
\end{align}
where $p_{\text{suc}} = \sum_{j \in A} \rho_{j,j}$ is the probability of the system having transitioned to a state labeled by the criteria as successfully merged. Assuming $p_1$ is nonzero, the state after the measurement is given by
\begin{align}
    \rho_1 &= \frac{1}{p_1} \Pi_A \rho_{\delta} \Pi_A  \nonumber\\
    &= \frac{1}{p_{\text{suc}}} \sum_{j,k \in A} \rho_{j,k} \op{\psi_j}{\psi_k}.
\end{align}
Thus, this measurement outcome is the desirable one: it heralds that the state has been projected onto the desired subspace. Furthermore, the probability of getting this outcome depends both on the overlap of the input state with the desired subspace and the parameter $\delta$, with values of $\delta \approx \frac{\pi}{2}$ maximizing the probability.

Consider now instead the '0' outcome. The probability for this outcome is
\begin{align}
    p_0 = 1 - p_1 = 1 - \sin(\delta)^2 p_{\text{suc}} 
\end{align}
and the post-measurement state is
\begin{align}
    \rho_0 &= \frac{1}{p_0} \Pi_B \rho_{\delta} \Pi_B \nonumber\\
    &= \frac{1}{p_0} \left( \cos(\delta)^2 \sum_{j,k \in A} \rho_{j,k} \op{\psi_j}{\psi_k} + \sum_{j,k \in B} \rho_{j,k} \op{\psi_j}{\psi_k} + \cos(\delta) \sum_{j \in A, k \in B} \left( \rho_{j,k} \op{\psi_j}{\psi_k} + \rho_{k,j} \op{\psi_k}{\psi_j} \right) \right) \; . \label{eq:post_0_oracle_state}
\end{align}
Note that this state closely resembles the input state from Eq.~\eqref{eq:input_to_oracle2}, except the components related to the $A$ subspace have decreased in magnitude by a factor of $\cos(\delta)$ and a renormalization by $1/p_0$ has occurred. To better understand this measurement-induced perturbation, we can define the following normalized states and coefficients
\begin{align}
    \rho_A = \rho_1 = \frac{1}{p_{\text{suc}}} \sum_{j,k \in A} \rho_{j,k} \op{\psi_j}{\psi_k} \quad
    \rho_B = \frac{1}{1-p_{\text{suc}}} \sum_{j,k \in B} \rho_{j,k} \op{\psi_j}{\psi_k}
\end{align}
\begin{align}   
    \Lambda_A(\delta, p_{\text{suc}}) &=  \frac{\sin(\delta)^2 (1-p_{\text{suc}}) \, p_{\text{suc}}}{(1-p_{\text{suc}}) + \cos(\delta)^2 \, p_{\text{suc}}} \\
    \Lambda_B(\delta, p_{\text{suc}}) &= \frac{\sin(\delta)^2 (1-p_{\text{suc}}) \, p_{\text{suc}}}{1- \sin(\delta)^2 \, p_{\text{suc}}}\\
    \Lambda_C(\delta, p_{\text{suc}}) &= \frac{1 - \cos(\delta) - \sin(\delta)^2 p_{\text{suc}}}{1 - \sin(\delta)^2 p_{\text{suc}} }
\end{align}
to rewrite the state as
\begin{align}
    \rho_0 =& \, \rho -  \Lambda_A(\delta, p_{\text{suc}}) \, \rho_A +  \Lambda_B(\delta, p_{\text{suc}}) \, \rho_B  - \Lambda_C(\delta, p_{\text{suc}}) \sum_{j \in A, k \in B} \left( \rho_{j,k} \op{\psi_j}{\psi_k} + \rho_{k,j} \op{\psi_k}{\psi_j} \right)\\
    \simeq& \, \, \rho - \delta^2 \left( 1 - p_{\text{suc}}\right) p_{\text{suc}} \, \rho_A + \delta^2 \left( 1 - p_{\text{suc}}\right) p_{\text{suc}} \, \rho_B \nonumber \\
    &-  \delta^2 \left( \frac{1}{2} - p_{\text{suc}} \right) \sum_{j \in A, k \in B} \left( \rho_{j,k} \op{\psi_j}{\psi_k} + \rho_{k,j} \op{\psi_k}{\psi_j} \right) + O(\delta^4) .
\end{align}
Thus, receiving a measurement result of '0' implies that the state has been left mostly unchanged, except primarily for two effects: The desired components related to $\rho_A$ have decreased by an amount $\Lambda_A \,\rho_A$, while the unwanted components related to $\rho_B$ have increased by an amount $\Lambda_B \,\rho_B$. In other words, the measurement has caused a shift towards the subspace of undesired states, with the magnitude of the shift depending on $\delta$ and $p_{\text{suc}}$.
From this, we see a trade-off in play when picking the parameter $\delta$.
As shown above, the largest probability of successful heralded projection requires $\delta \approxeq \frac{\pi}{2}$, but in this case the desired $A$ part of the state is completely lost whenever the measurement outcome `0' occurs (see Eq.~\eqref{eq:post_0_oracle_state}). 
On the other hand, a small value of $\delta$ implies a small probability of successful heralded projection but also a small state perturbation in the case of the outcome `0', with both scaling as $\delta^2$ in the small-$\delta$ limit. 
In this sense, $\delta$ represents the power of the measurement, with strong measurements yielding a higher chance of detecting a successful molecular merger but also a higher disruptive impact of the measurement on the state.
Picking good schedules for adjusting $\delta$ has previously been studied in the context of Grover search~\cite{andres2022weakly,yan2022fixed,mizel2009critically}, and will likely also play a significant role in determining the performance of the simulation approaches presented here.

\end{document}